\documentclass[12pt,letterpaper]{article}
\usepackage{epsfig,rotating,setspace,latexsym,amsmath,epsf,amssymb,amsfonts,bm,theorem,cite,caption,subcaption,enumerate,longtable,accents}
\usepackage{algorithm,algorithmic,graphicx,epsf,authblk,epstopdf,url,color,multirow}
\newtheorem{theorem}{Theorem}

\newtheorem{corollary}{Corollary}

\newtheorem{lemma}{Lemma}
\newenvironment{Proof}[1]{\medskip\par\noindent{\bf Proof:\,}\,#1}{{\mbox{\,$\blacksquare$}\par}}

\newcommand{\bg}{{\mathbf{G}}}
\newcommand{\bs}{{\mathbf{S}}}
\newcommand{\cq}{{\mathcal{Q}}}
\newcommand{\cp}{{\mathcal{P}}}

\allowdisplaybreaks

\begin{document}
	
\title{Multi-Message Private Information Retrieval: Capacity Results and Near-Optimal Schemes\thanks{This work was supported by NSF Grants CNS 13-14733, CCF 14-22111, CCF 14-22129, and CNS 15-26608. A shorter version is submitted to IEEE ISIT 2017.}}
	
\author{Karim Banawan \qquad Sennur Ulukus\\
	\normalsize Department of Electrical and Computer Engineering\\
	\normalsize University of Maryland, College Park, MD 20742 \\
	\normalsize {\it kbanawan@umd.edu} \qquad {\it ulukus@umd.edu}}
	
\maketitle
	
\vspace*{-0.5cm}

\begin{abstract}	
We consider the problem of multi-message private information retrieval (MPIR) from $N$ non-communicating replicated databases. In MPIR, the user is interested in retrieving $P$ messages out of $M$ stored messages without leaking the identity of the retrieved messages. The information-theoretic sum capacity of MPIR $C_s^P$ is the maximum number of desired message symbols that can be retrieved privately per downloaded symbol. For the case $P \geq \frac{M}{2}$, we determine the exact sum capacity of MPIR as $C_s^P=\frac{1}{1+\frac{M-P}{PN}}$. The achievable scheme in this case is based on downloading MDS-coded mixtures of all messages. For $P \leq \frac{M}{2}$, we develop lower and upper bounds for all $M,P,N$. These bounds match if the total number of messages $M$ is an integer multiple of the number of desired messages $P$, i.e., $\frac{M}{P} \in \mathbb{N}$. In this case, $C_s^P=\frac{1-\frac{1}{N}}{1-(\frac{1}{N})^{M/P}}$. The achievable scheme in this case generalizes the single-message capacity achieving scheme to have unbalanced number of stages per round of download. For all the remaining cases, the difference between the lower and upper bound is at most $0.0082$, which occurs for $M=5$, $P=2$, $N=2$. Our results indicate that joint retrieval of desired messages is more efficient than successive use of single-message retrieval schemes.
\end{abstract}
	
\section{Introduction}

The privacy of the contents of the downloaded information from curious public databases has attracted considerable research within the computer science community \cite{ChorPIR, yekhanin2010private, PIRsurvey2004, ostrovsky2007survey}. The problem is motivated by practical examples such as: ensuring privacy of investors as they download records in a stock market, since revealing the interest in a certain record may influence its value; ensuring the privacy of an inventor as they look up existing patents in a database, since revealing what they are looking at leaks some information about the current invention they are working on; and protecting activists in oppressive regimes as they browse restricted content on the internet \cite{yekhanin2010private}. In the classical private information retrieval (PIR) problem, a user wishes to download a certain message (or file) from $N$ non-communicating databases without leaking any information about the identity of the downloaded message. The contents of the databases are identical. The user performs this operation by preparing and submitting queries to all databases. The databases respond truthfully with answer strings which are functions of the queries and the messages. The user needs to reconstruct the desired message from these answer strings. A trivial solution for this seemingly difficult problem is for the user to download the contents of all databases. This solution however is extremely inefficient. The efficiency is measured by the retrieval rate which is the ratio of the number of retrieved desired message symbols to the number of total downloaded symbols. The capacity of PIR is the maximum retrieval rate over all possible PIR schemes.

The computer science formulation of this problem assumes that the messages are of length one. The metrics in this case are the download cost, i.e., the sum of lengths of the answer strings, and the upload cost, i.e., the size of the queries. Most of this work is computational PIR as it ensures only that a server cannot get any information about user intent unless it solves a certain computationally hard problem \cite{yekhanin2010private, cachin1999computationally}. The information-theoretic re-formulation of the problem considers arbitrarily large message sizes, and ignores the upload cost. This formulation provides an absolute, i.e., information-theoretic, guarantee that no server participating in the protocol gets any information about the user intent. Towards that end, recently, \cite{JafarPIRBlind} has drawn a connection between the PIR problem and the blind interference alignment scheme proposed in \cite{BIA}. Then, \cite{JafarPIR} has determined the exact capacity of the classical PIR problem. The retrieval scheme in \cite{JafarPIR} is based on three principles: message symmetry, symmetry across databases, and exploiting side information from the undesired messages through alignment.

The basic PIR setting has been extended in several interesting directions. The first extension is the coded PIR (CPIR) problem \cite{RamchandranPIR,RazanPIR,YamamotoPIR}. The contents of the databases in this problem are coded by an $(N,K)$ storage code instead of being replicated. This is a natural extension since most storage systems nowadays are in fact coded to achieve reliability against node failures and erasures with manageable storage cost. In \cite{KarimCoded}, the exact capacity of the MDS-coded PIR is determined. Another interesting extension is PIR with colluding databases (TPIR). In this setting, $T$ databases can communicate and exchange the queries to identify the desired message. The exact capacity of colluded PIR is determined in \cite{JafarColluding}. The case of coded colluded PIR is investigated in \cite{codedcolluded}. The robust PIR problem (RPIR) extension considers the case when some databases are not responsive \cite{JafarColluding}. Lastly, in the symmetric PIR problem (SPIR) the privacy of the remaining records should be maintained against the user in addition to the usual privacy constraint on the databases, i.e., the user should not learn any other messages other than the one it wished to retrieve. The exact capacity of symmetric PIR is determined in \cite{symmetricPIR}; and the exact capacity of symmetric PIR from coded databases is determined in \cite{codedsymmetric}.

In some applications, the user may be interested in retrieving multiple messages from the databases without revealing the identities of these messages. Returning to the examples presented earlier: the investor may be interested in comparing the values of multiple records at the same time, and the inventor may be looking up several patents that are closely related to their work. One possible solution to this problem is to use single-message retrieval scheme in \cite{JafarPIR} successively. We show in this work that multiple messages can be retrieved more efficiently than retrieving them one-by-one in a sequence. This resembles superiority of joint decoding in multiple access channels over multiple simultaneous single-user transmissions \cite{cover}. To motivate the multi-message private information retrieval problem (MPIR), consider the example in \cite[Section~4.3]{JafarPIR} where the number of messages is $M=3$, number of databases is $N=2$, and the user is interested in retrieving only $P=1$ message. Here the optimal scheme retrieves 8 desired bits in 14 downloads, hence with a rate $4/7$. When the user wishes to retrieve $P=2$ messages, if we use the scheme in \cite{JafarPIR} twice in a row, we retrieve 16 bits in 28 downloads, hence again a \emph{sum rate} of $4/7$. Even considering the fact that the scheme in \cite{JafarPIR} retrieves 2 bits of the second message \emph{for free} in downloading the first message, i.e., it actually retrieves 10 bits in 14 downloads, hence a sum rate of $5/7$, we show in this paper that a better sum rate of $4/5$ can be achieved by joint retrieval of the messages.

Although there is a vast literature on classical PIR in the computer science literature, only a few works exist in MPIR, such as:  \cite{OneblockFits} which proposes a multi-block (multi-message) scheme and observes that if the user requests multiple blocks (messages), it is possible to reuse randomly mixed data blocks (answer strings) across multiple requests (queries). \cite{RAID} develops a multi-block scheme which further reduces the communication overhead. \cite{multiblock} develops an achievable scheme for the multi-block PIR by designing $k$-safe binary matrices that uses XOR operations. \cite{multiblock} extends the scheme in \cite{ChorPIR} to multiple blocks. \cite{multiquery} designs an efficient non-trivial multi-query computational PIR protocol and gives a lower bound on the communication of any multi-query information retrieval protocol. These works do not consider determining the information-theoretic capacity.

In this paper, we formulate the MPIR problem with non-colluding repeated databases from an information-theoretic perspective. Our goal is to characterize the sum capacity of the MPIR problem $C_s^P$, which is defined as the maximum ratio of the number of retrieved symbols from the $P$ desired messages to the number of total downloaded symbols. When the number of desired messages $P$ is at least half of the total number of messages $M$, i.e., $P \geq \frac{M}{2}$, we determine the exact sum capacity of MPIR as $C_s^P=\frac{1}{1+\frac{M-P}{PN}}$. We use a novel achievable scheme which downloads MDS-coded mixtures of all messages. We show that joint retrieving of the desired messages strictly outperforms successive use of single-message retrieval for $P$ times. Additionally, we present an achievable rate region to characterize the trade-off between the retrieval rates of the desired $P$ messages.

For the case of $P \leq \frac{M}{2}$, we derive lower and upper bounds that match if the total number of messages $M$ is an integer multiple of the number of desired messages $P$, i.e., $\frac{M}{P} \in \mathbb{N}$. In this case, the sum capacity is $C_s^P=\frac{1-\frac{1}{N}}{1-(\frac{1}{N})^{M/P}}$. The result resembles the single-message capacity with the number of messages equal to $\frac{M}{P}$. In other cases, although the exact capacity is still an open problem, we show numerically that the gap between the lower and upper bounds is monotonically decreasing in $N$ and is upper bounded by $0.0082$. The achievable scheme when $P \leq \frac{M}{2}$ is inspired by the greedy algorithm in \cite{JafarPIR}, which retrieves all possible combinations of messages. The main difference of our scheme from the scheme in \cite{JafarPIR} is the number of stages required in each download round. For example, round $M-P+1$ to round $M-1$, which correspond to retrieving the sum of $M-P+1$ to sum of $M-1$ messages, respectively, are suppressed in our scheme. This is because, they do not generate any useful side information for our purposes here, in contrast to \cite{JafarPIR}. Interestingly, the number of stages for each round is related to the output of a $P$-order IIR filter \cite{oppenheim}. Our converse proof generalizes the proof in \cite{JafarPIR} for $P \geq 1$. The essence of the proof is captured in two lemmas: the first lemma lower bounds the uncertainty of the interference for the case $P \geq \frac{M}{2}$, and the second lemma upper bounds the remaining uncertainty after conditioning on $P$ interfering messages.

\section{Problem Formulation}

Consider a classical PIR setting storing $M$ messages (or files). Each message is a vector $W_i \in \mathbb{F}_q^L,\: i \in \{1, \cdots, M\} $, whose elements are picked uniformly and independently from sufficiently large field $\mathbb{F}_q$. Denote the contents of message $W_m$ by the vector $[w_m(1), w_m(2), \cdots, w_m(L)]^T$. The messages are independent and identically distributed, and thus,
\begin{align}
H(W_i)&=L, \quad i \in \{1, \cdots, M\} \\
\label{msg_indep}H\left(W_{1:M}\right)&=ML
\end{align}
where $W_{1:M}=(W_1,W_2, \cdots, W_M)$. The messages are stored in $N$ non-colluding (non-communicating) databases. Each database stores an identical copy of all $M$ messages, i.e., the databases encode the messages via $(N,1)$ repetition storage code \cite{KarimCoded}.

In the MPIR problem, the user aims to retrieve a subset of messages indexed by the index set $\mathcal{P}=\{i_1, \cdots, i_P\} \subseteq \{1, \cdots, M\}$ out of the available messages, where $|\mathcal{P}|=P$, without leaking the identity of the subset $\mathcal{P}$. We assume that the cardinality of the potential message set, $P$, is known to all databases. To retrieve $W_\mathcal{P}=(W_{i_1}, W_{i_2}, \cdots, W_{i_P})$, the user generates a query $Q_n^{[\mathcal{P}]}$ and sends it to the $n$th database. The user does not have any knowledge about the messages in advance, hence the messages and the queries are statistically independent,
\begin{align}\label{msg_query_indep}
I\left(W_1, \cdots, W_M;Q_1^{[\mathcal{P}]}, \cdots, Q_N^{[\mathcal{P}]} \right)=I\left(W_{1:M};Q_{1:N}^{[\mathcal{P}]}\right)=0
\end{align}
The privacy is satisfied by ensuring statistical independence between the queries and the message index set $\mathcal{P}=\{i_1, \cdots, i_P\}$, i.e., the privacy constraint is given by,
\begin{align}\label{privacy_constraint}
I\left(Q_n^{[i_1, \cdots, i_P]};i_1, \cdots, i_P\right)=I\left(Q_n^{[\mathcal{P}]};\mathcal{P}\right)=0, \quad n \in \{1, \cdots, N\}
\end{align}
The $n$th database responds with an answer string $A_n^{[\mathcal{P}]}$, which is a deterministic function of the queries and the messages, hence
\begin{align}
H(A_n^{[\mathcal{P}]}|Q_n^{[\mathcal{P}]},W_{1:M})=0
\end{align}
We further note that by the data processing inequality and (\ref{privacy_constraint}),
\begin{align}
I\left(A_n^{[\mathcal{P}]};\mathcal{P}\right)=0, \quad n \in \{1, \cdots, N\}
\end{align}
In addition, the user should be able to reconstruct the messages $W_\mathcal{P}$ reliably from the collected answers from all databases given the knowledge of the queries. Thus, we write the reliability constraint as,
\begin{align}\label{reliability}
H\left(W_{i_1}, \cdots, W_{i_P}|A_1^{[\mathcal{P}]}, \cdots, A_N^{[\mathcal{P}]}, Q_1^{[\mathcal{P}]}, \cdots, Q_N^{[\mathcal{P}]}\right)=H\left(W_\mathcal{P}|A_{1:N}^{[\mathcal{P}]},Q_{1:N}^{[\mathcal{P}]}\right)=0
\end{align}

We denote the retrieval rate of the $i$th message by $R_i$, where $i \in \mathcal{P}$. The retrieval rate of the $i$th message is the ratio between the length of message $i$ and the total download cost of the message set $\mathcal{P}$ that includes $W_i$. Hence,
\begin{align}
R_i=\frac{H(W_i)}{\sum_{n=1}^N H\left(A_n^{[\mathcal{P}]}\right)}
\end{align}
The sum retrieval rate of $W_\mathcal{P}$ is given by,
\begin{align}
\sum_{i=1}^PR_i=\frac{H(W_\mathcal{P})}{\sum_{n=1}^N H\left(A_n^{[\mathcal{P}]}\right)}=\frac{PL}{\sum_{n=1}^N H\left(A_n^{[\mathcal{P}]}\right)}
\end{align}
The sum capacity of the MPIR problem is given by
\begin{align}
C_s^P=\sup \: \sum_{i=1}^PR_i
\end{align}
where the $\sup$ is over all private retrieval schemes.

In this paper, we follow the information-theoretic assumptions of large enough message size, large enough field size, and ignore the upload cost as  in \cite{KarimCoded,JafarPIR,JafarColluding,YamamotoPIR}. A formal treatment of the capacity under message and field size constraints for $P=1$ can be found in \cite{arbmsgPIR}. We note that the MPIR problem described here reduces to the classical PIR problem when $P=1$, whose capacity is characterized in \cite{JafarPIR}.

\section{Main Results and Discussions}

Our first result is the exact characterization of the sum capacity for the case $P \geq \frac{M}{2}$, i.e., when the user wishes to privately retrieve at least half of the messages stored in the databases.
\begin{theorem}\label{thm1}
For the MPIR problem with non-colluding and replicated databases, if the number of desired messages $P$ is at least half of the number of overall stored messages $M$, i.e., if $P \geq \frac{M}{2}$, then the sum capacity is given by,
\begin{align}\label{result1}
C_s^P=\frac{1}{1+\frac{M-P}{PN}}
\end{align}
\end{theorem}

The achievability proof for Theorem~\ref{thm1} is given in Section~\ref{sec:achievable1}, and the converse proof is given in Section~\ref{converse1}. We note that when $P=1$, the constraint of Theorem~\ref{thm1} is equivalent to $M=2$, and the result in  (\ref{result1}) reduces to the known result of \cite{JafarPIR} for $P=1$, $M=2$, which is $\frac{1}{1+\frac{1}{N}}$. We observe that the sum capacity in (\ref{result1}) is a strictly increasing function of $N$, and $C_s^P \rightarrow 1$ as $N \rightarrow \infty$. We also observe that the sum capacity in this regime is a strictly increasing function of $P$, and approaches $1$ as $P \rightarrow M$.

The following corollary compares our result and the rate corresponding to the repeated use of single-message retrieval scheme \cite{JafarPIR}.
\begin{corollary}\label{corrollary1}
For the MPIR problem with $P \geq \frac{M}{2}$, the repetition of the single-message retrieval scheme of \cite{JafarPIR} $P$ times in a row, which achieves a sum rate of,
\begin{align}
R_s^{rep}=\frac{(N-1)(N^{M-1}+P-1)}{N^M-1}
\label{repeat}
\end{align}
is strictly sub-optimal with respect to the exact capacity in (\ref{result1}).
\end{corollary}

\begin{Proof}
In order to use the single-message capacity achieving PIR scheme as an MPIR scheme, the user repeats the single-message achievable scheme for each individual message that belongs to $\cp$. We note that at each repetition, the scheme downloads extra decodable symbols from other messages. By this argument, the following rate $R_s^{rep}$ is achievable using a repetition of the single-message scheme,
\begin{align}
R_s^{rep}=C+\Delta(M,P,N) \label{eqn-delta}
\end{align}
where $C$ is the single-message capacity which is given by $C=\frac{1-\frac{1}{N}}{1-(\frac{1}{N})^M}$ \cite{JafarPIR}, and $\Delta(M,P,N)$ is the rate of the extra decodable symbols that belong to $\cp$. To calculate $\Delta(M,P,N)$, we note that the total download cost $D$ is given by $D=\frac{L}{C}$ by definition. Since $L=N^M$ in the single-message scheme, $D=\frac{N^M(1-(\frac{1}{N})^M)}{1-\frac{1}{N}}=\frac{N^{M+1}-N}{N-1}$. The single-message scheme downloads one symbol from every message from every database, i.e., the scheme downloads extra $(P-1)N$ symbols from the remaining desired messages that belong to $\cp$, thus,
\begin{align}
\Delta(M,P,N)=\frac{(P-1)N(N-1)}{N^{M+1}-N}=\frac{(P-1)(N-1)}{N^M-1}
\end{align}
Using this in (\ref{eqn-delta}) gives the $R_s^{rep}$ expression in (\ref{repeat}).

Now, the difference between the capacity in (\ref{result1}) and achievable rate in (\ref{repeat}) is,
\begin{align}
C_s^P-R_s^{rep}&=\frac{PN}{P(N-1)+M}-\frac{(N-1)(N^{M-1}+P-1)}{N^M-1}\\
& =\frac{\eta(P,M,N)}{(N^M-1)(P(N-1)+M)}
\end{align}
It suffices to prove that $\eta(P,M,N) \geq 0$ for all $P$, $M$, $N$ when $P \geq \frac{M}{2}$ and $N \geq 2$. Note,
\begin{align}\label{corr0}
\eta(P,M,N)=&(2P-M)N^M+(M-P)N^{M-1} -P(P-1)N^2\notag\\
            &+((P-1)(2P-M)-P)N+(M-P)(P-1)
\end{align}
In the regime $P \geq \frac{M}{2}$, coefficients of $N^M,N^{M-1},N^0$ are non-negative. Denote the negative terms in $\eta(\cdot)$ by $\nu(P,N)$ which is $\nu(P,N)=P(P-1)N^2+PN$. We note $\nu(P,N) < P^2N^2$ when $N>1$, which is the case here. Thus,
\begin{align}
\eta(P,M,N) \geq &(2P-M)N^M+(M-P)N^{M-1} \notag\\
&+(P-1)(2P-M)N+(M-P)(P-1)-P^2N^2\\
> &(2P-M)N^M+(M-P)N^{M-1}-P^2N^2\\
= &N^2\left((2P-M)N^{M-2}+(M-P)N^{M-3}-P^2\right)\\
\label{corr1}\geq & N^2\left((2P-M)2^{M-2}+(M-P)2^{M-3}-P^2\right)\\
=& N^2\left(2^{M-3}(3P-M)-P^2\right) \\
\label{corr2} \geq & N^2\left(2^{M-3}\cdot\frac{M}{2}-M^2\right)\\
\label{corr3} =&MN^2\left(2^{M-4}-M\right)
\end{align}
where (\ref{corr1}) follows from the fact that $(2P-M)N^{M-2}+(M-P)N^{M-3}-P^2$ is monotone increasing in $N \geq 2$ for $M\geq 3$, and (\ref{corr2}) follows from $\frac{M}{2} \leq P \leq M$. From (\ref{corr3}), we conclude that $\eta(M,P,N) > 0$ for all $M \geq 7$, $P \geq \frac{M}{2}$ and $N \geq 2$. Examining the expression in (\ref{corr0}) for the remaining cases manually, i.e., when $M \leq 6$, we note that $\eta(M,P,N) > 0$ in these cases as well. Therefore, $\eta(M,P,N) > 0$ for all possible cases, and the MPIR capacity is strictly larger than the rate achieved by repeating the optimum single-message PIR scheme.
\end{Proof}

For the example in the introduction, where $M=3$, $P=2$, $N=2$, our MPIR scheme achieves a sum capacity of $\frac{4}{5}$ in (\ref{result1}), which is strictly larger than the repeating-based achievable sum rate of $\frac{5}{7}$ in (\ref{repeat}).

The following corollary gives an achievable rate region for the MPIR problem.

\begin{corollary}\label{corollary2}
For the MPIR problem, for the case $P \geq \frac{M}{2}$, the following rate region is achievable,
\begin{align}\label{achievable_region}
\mathcal{C}=\emph{conv} & \left\{ (C,\delta, \cdots, \delta), (\delta, C, \cdots, \delta), \cdots, (\delta, \cdots, \delta, C) , (C,0,0,\cdots,0),\right.\notag\\
&\:\:\left. (0,C,0,\cdots,0), \cdots, (0,0,\cdots,C), (0,0, \cdots, 0), \left(C^P,C^P, \cdots, C^P\right) \right\}
\end{align}
where
\begin{align}
C=\frac{1-\frac{1}{N}}{1-(\frac{1}{N})^M}, \qquad
C^P=\frac{C_s^P}{P}=\frac{N}{PN+(M-P)}, \qquad
\delta=\frac{\Delta(M,P,N)}{P-1}=\frac{N-1}{N^M-1}
\end{align}
and where $\emph{conv}$ denotes the convex hull, and all corner points lie in the $P$-dimensional space.
\end{corollary}

\begin{Proof}
This is a direct consequence of Theorem~\ref{thm1} and Corollary~\ref{corrollary1}. The corner point  $\left(C,\frac{\Delta(M,P,N)}{P-1},\frac{\Delta(M,P,N)}{P-1}, \cdots, \frac{\Delta(M,P,N)}{P-1}\right)=\left(\frac{1-\frac{1}{N}}{1-(\frac{1}{N})^M}, \frac{N-1}{N^M-1},\frac{N-1}{N^M-1}, \cdots, \frac{N-1}{N^M-1}\right)$ is achievable from the single-message achievable scheme. Due to the symmetry of the problem any other permutation for the coordinates of this corner point is also achievable by changing the roles of the desired messages. Theorem~\ref{thm1} gives the symmetric sum capacity corner point for the case of $P \geq \frac{M}{2}$, namely $\left(\frac{C_s^P}{P}, \frac{C_s^P}{P}, \cdots, \frac{C_s^P}{P}\right) =\left(\frac{N}{PN+(M-P)}, \frac{N}{PN+(M-P)}, \cdots, \frac{N}{PN+(M-P)}\right)$. By time sharing of these corner points along with the origin, the  region in (\ref{achievable_region}) is achievable.
\end{Proof}

\begin{figure}[t]
\centering
\includegraphics[width=0.4\textwidth]{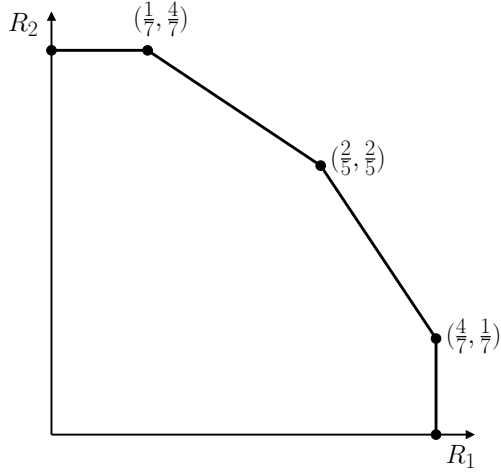}
\caption{The achievable rate region of $M=3$, $P=2$, $N=2$.}
\label{fig:region322}
\end{figure}

As an example for this achievable region, consider again the example in the introduction, where $M=3$, $P=2$, $N=2$. In this case, we have a two-dimensional rate region with three corner points: $(\frac{4}{7},\frac{1}{7})$, which corresponds to the single-message capacity achieving point that aims at retrieving $W_1$;  $(\frac{1}{7},\frac{4}{7})$, which corresponds to single-message capacity achieving point that aims at retrieving $W_2$; and $(\frac{2}{5},\frac{2}{5})$, which corresponds to the symmetric sum capacity point. The convex hull of these corner points together with the points on the axes gives the achievable region in Fig.~\ref{fig:region322}.

For the case $P \leq \frac{M}{2}$, we have the following result, where the lower and upper bound match if $\frac{M}{P} \in \mathbb{N}$.

\begin{theorem}\label{thm2}
For the MPIR problem with non-colluding and replicated databases, when $P \leq \frac{M}{2}$, the sum capacity is lower and upper bounded as,
\begin{align}\label{bounds}
\underaccent{\bar}{R}_s \leq C_s^P \leq \bar{R}_s
\end{align}
where the upper bound $\bar{R}_s$ is given by,
\begin{align}\label{upper_bound}
\bar{R}_s &=\frac{1}{1+\frac{1}{N}+\cdots+\frac{1}{N^{\lfloor\frac{M}{P}\rfloor-1}}+\left(\frac{M}{P}-\lfloor\frac{M}{P}\rfloor\right)\frac{1}{N^{\lfloor \frac{M}{P}\rfloor}}} \\ &=\frac{1}{\frac{1-(\frac{1}{N})^{\lfloor\frac{M}{P}\rfloor}}{1-\frac{1}{N}}+\left(\frac{M}{P}-\left\lfloor\frac{M}{P}\right\rfloor\right)\frac{1}{N^{\left\lfloor \frac{M}{P}\right\rfloor}}}
\end{align}
For the lower bound, define $r_i$ as,
\begin{align}\label{root}
r_i=\frac{e^{j2\pi(i-1)/P}}{N^{1/P}-e^{j2\pi(i-1)/P}}, \quad i=1,\cdots, P
\end{align}
where $j=\sqrt{-1}$, and denote $\gamma_i, \: i=1, \cdots, P$, to be the solutions of the linear equations $\sum_{i=1}^P \gamma_i r_i^{-P}=(N-1)^{M-P}$, and $\sum_{i=1}^P \gamma_i r_i^{-k}=0, \: k=1, \cdots, P-1$, then $\underaccent{\bar}{R}_s$ is given by,
\begin{align} \label{ach-rate}
\underaccent{\bar}{R}_s=\frac{\sum_{i=1}^{P}\gamma_i r_i^{M-P}\left[\left(1+\frac{1}{r_i}\right)^M-\left(1+\frac{1}{r_i}\right)^{M-P}\right]}{\sum_{i=1}^{P}\gamma_i r_i^{M-P}\left[\left(1+\frac{1}{r_i}\right)^M-1\right]}
\end{align}
\end{theorem}

The achievability lower bound in Theorem~\ref{thm2} is shown in Section~\ref{sec:achievable2} and the upper bound is derived in Section~\ref{converse2}. The following corollary states that the bounds in Theorem~\ref{thm2} match if the total number of messages is an integer multiple of the number of desired messages.

\begin{corollary}\label{corollary3}
For the MPIR problem  with non-colluding and replicated databases, if $\frac{M}{P}$ is an integer, then the bounds in (\ref{bounds}) match, and hence,
\begin{align}
C_s^P=\frac{1-\frac{1}{N}}{1-(\frac{1}{N})^{\frac{M}{P}}}, \quad \frac{M}{P} \in \mathbb{N}
\end{align}
\end{corollary}

\begin{Proof}
For the upper bound, observe that if $\frac{M}{P} \in \mathbb{N}$, then $\frac{M}{P}=\left\lfloor\frac{M}{P}\right\rfloor$. Hence, (\ref{upper_bound}) becomes
\begin{align}
\bar{R}_s=\frac{1-\frac{1}{N}}{1-(\frac{1}{N})^{\frac{M}{P}}}
\end{align}
For the lower bound, consider the case $\frac{M}{P} \in \mathbb{N}$. From (\ref{root}),
\begin{align}
\left(1+\frac{1}{r_i}\right)^M=\left(\frac{N^{1/P}}{e^{j2\pi(i-1)/P}}\right)^M=N^{\frac{M}{P}}
\end{align}
since $e^{j2\pi(i-1)M/P}=1$ for $\frac{M}{P} \in \mathbb{N}$. Similarly, $\left(1+\frac{1}{r_i}\right)^{M-P}=N^{\frac{M}{P}-1}$. Hence, if  $\frac{M}{P} \in \mathbb{N}$,
\begin{align}
\underaccent{\bar}{R}_s&=\frac{\sum_{i=1}^{P}\gamma_i r_i^{M-P}\left[N^{\frac{M}{P}}-N^{\frac{M}{P}-1}\right]}{\sum_{i=1}^{P}\gamma_i r_i^{M-P}\left[N^{\frac{M}{P}}-1\right]}\\
&=\frac{N^{\frac{M}{P}}-N^{\frac{M}{P}-1}}{N^{\frac{M}{P}}-1} \\
&=\frac{1-\frac{1}{N}}{1-(\frac{1}{N})^{\frac{M}{P}}}
\end{align}
Thus, $\underaccent{\bar}{R}_s=C_s^P=\bar{R}_s$ if $\frac{M}{P} \in \mathbb{N}$, and we have an exact capacity result in this case.
\end{Proof}

Examining the result, we observe that when the total number of messages is an integer multiple of the number of  desired messages, the sum capacity of the MPIR is the same as the capacity of the single-message PIR with the number of messages equal to $\frac{M}{P}$. Note that, although at first the result may seem as if every $P$ messages can be lumped together as a single message, and the achievable scheme in \cite{JafarPIR} can be used, this is not the case. The reason for this is that, we need to ensure the privacy constraint for \emph{every subset} of messages of size $P$. That is why, in this paper, we develop a new achievable scheme.

\begin{figure}[t]
\centering
\includegraphics[width=0.5\textwidth]{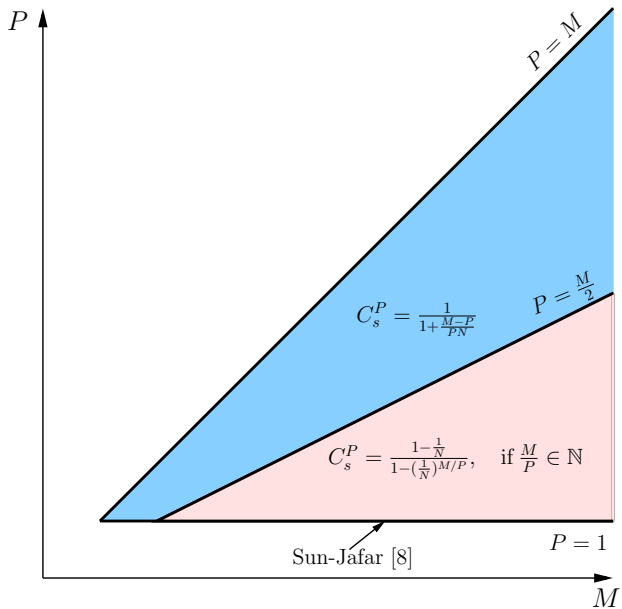}
\caption{Summary of the state of the results.}
\label{fig:main_result}
\end{figure}

The state of the results is summarized in Fig.~\ref{fig:main_result}: Consider the $(M,P)$ plane, where naturally $M\geq P$. The valid part of the plane is divided into two regions. The first region is confined between the lines $P=\frac{M}{2}$ and $P=M$; the sum capacity in this region is exactly characterized (Theorem~\ref{thm1}). The second region is confined between the lines $P=1$ and $P=\frac{M}{2}$; the sum capacity in this region is characterized only for the cases when $\frac{M}{P} \in \mathbb{N}$ (Corollary~\ref{corollary3}). The line $P=1$ corresponds to the previously known result for the single-message PIR \cite{JafarPIR}. The exact capacity for the rest of the cases is still an open problem; however, the achievable scheme in Theorem~\ref{thm2} yields near-optimal sum rates for all the remaining cases with the largest difference of $0.0082$ from the upper bound, as discussed next.

\begin{figure}[t]
\centering
\begin{subfigure}[b]{0.48\textwidth}
\includegraphics[width=\textwidth]{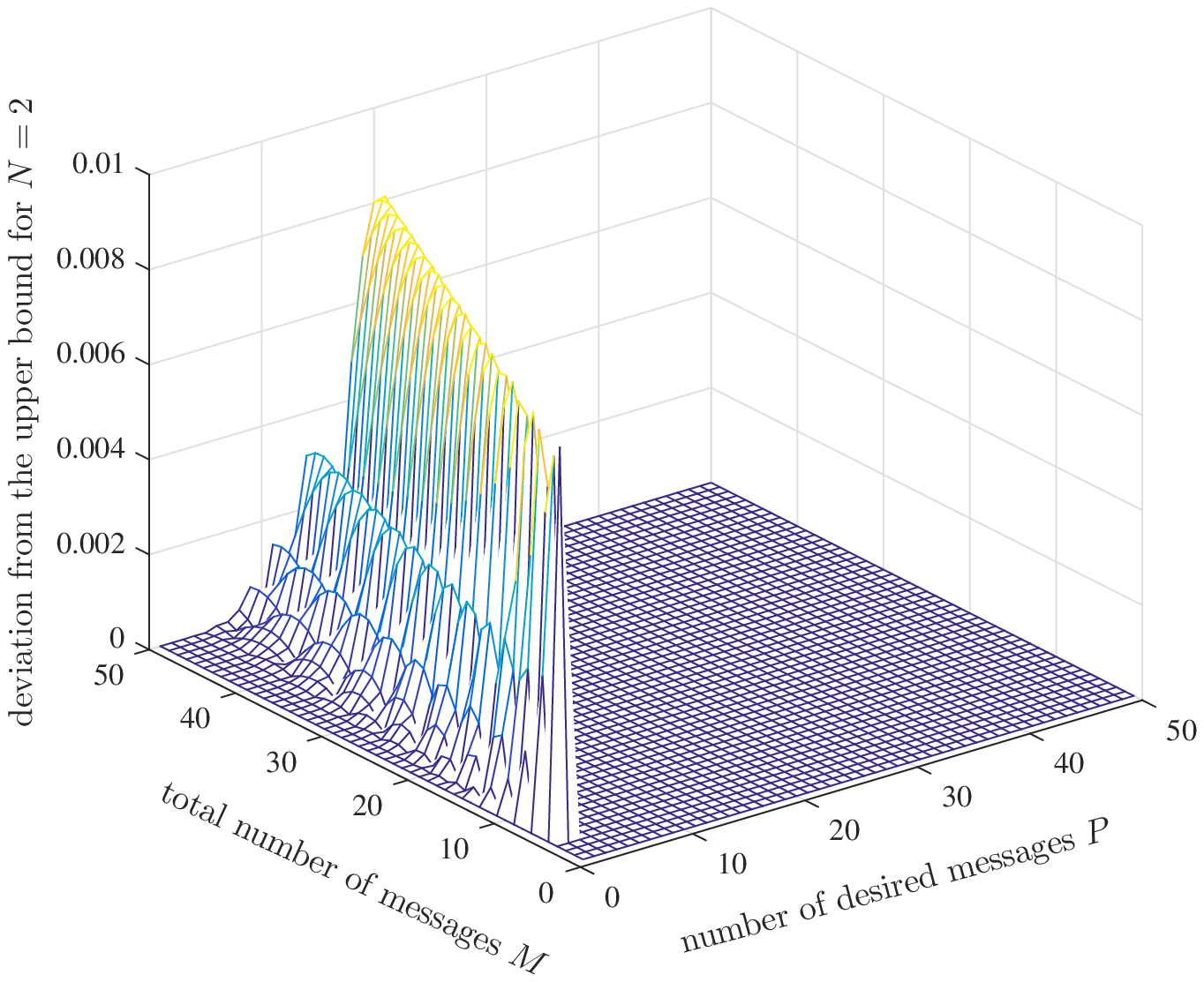}
\caption{$N=2$}
\end{subfigure}
\begin{subfigure}[b]{0.48\textwidth}
\includegraphics[width=\textwidth]{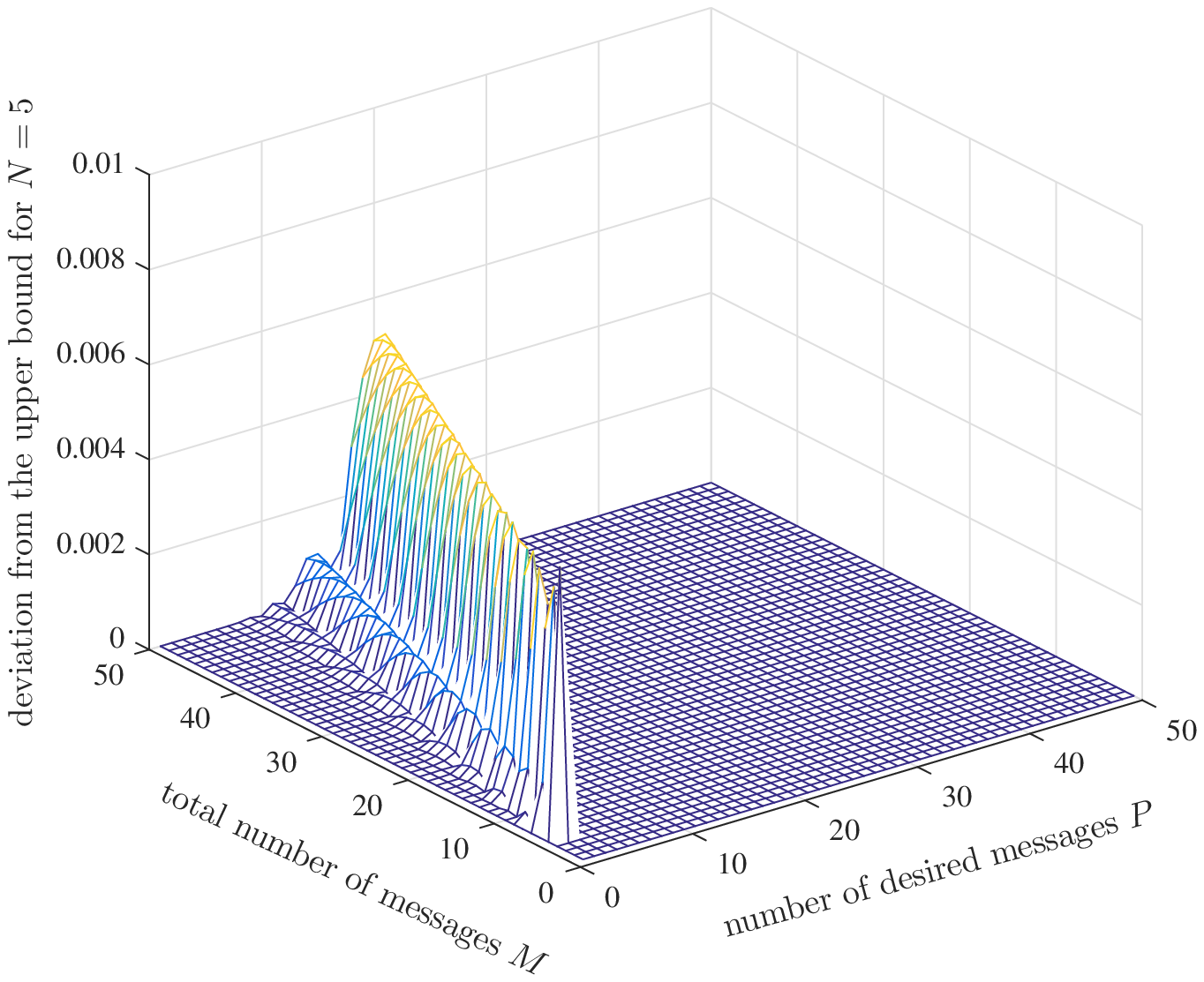}
\caption{$N=5$}
\end{subfigure}
\begin{subfigure}[b]{0.48\textwidth}
\includegraphics[width=\textwidth]{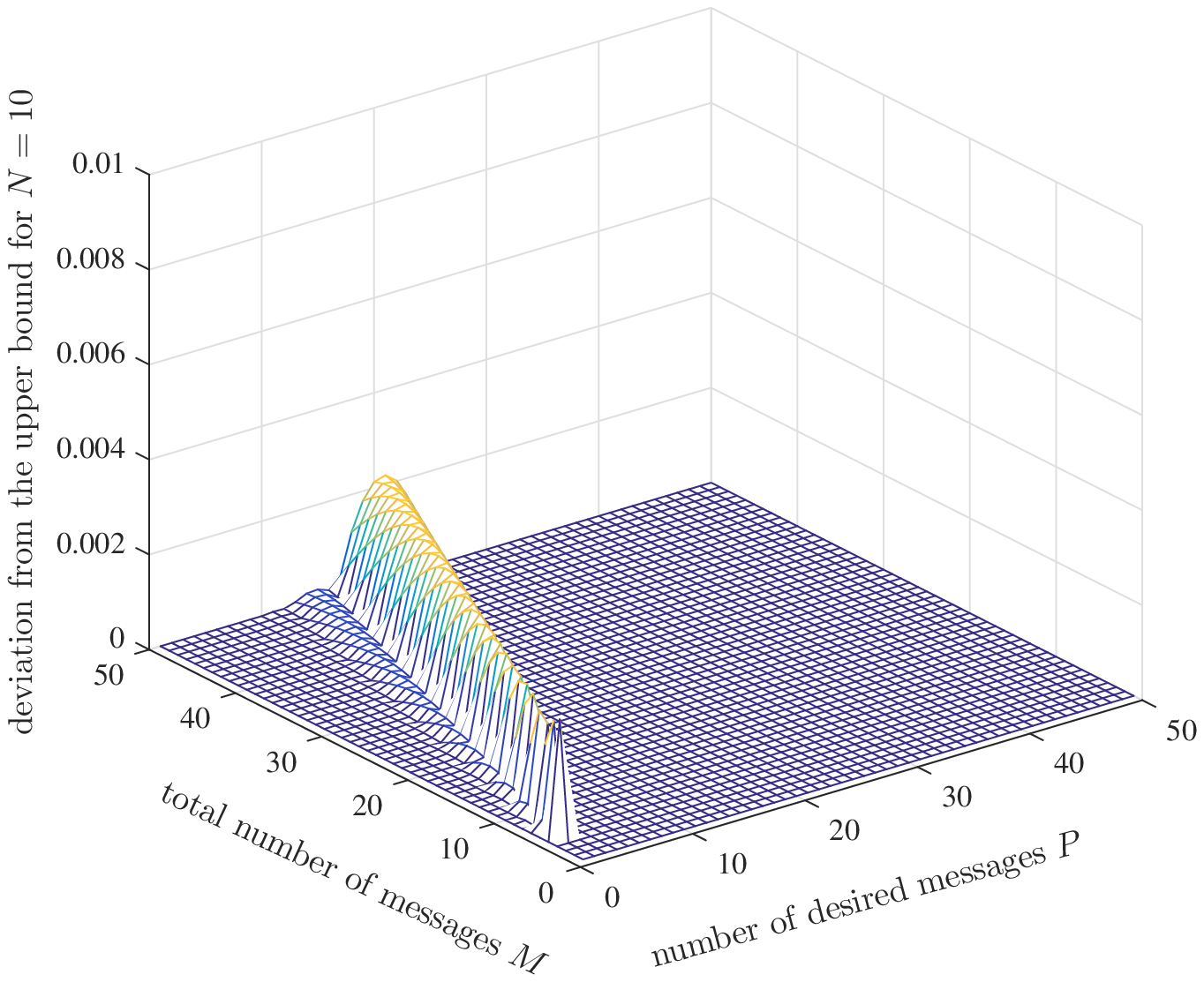}
\caption{$N=10$}
\end{subfigure}
\begin{subfigure}[b]{0.48\textwidth}
\includegraphics[width=\textwidth]{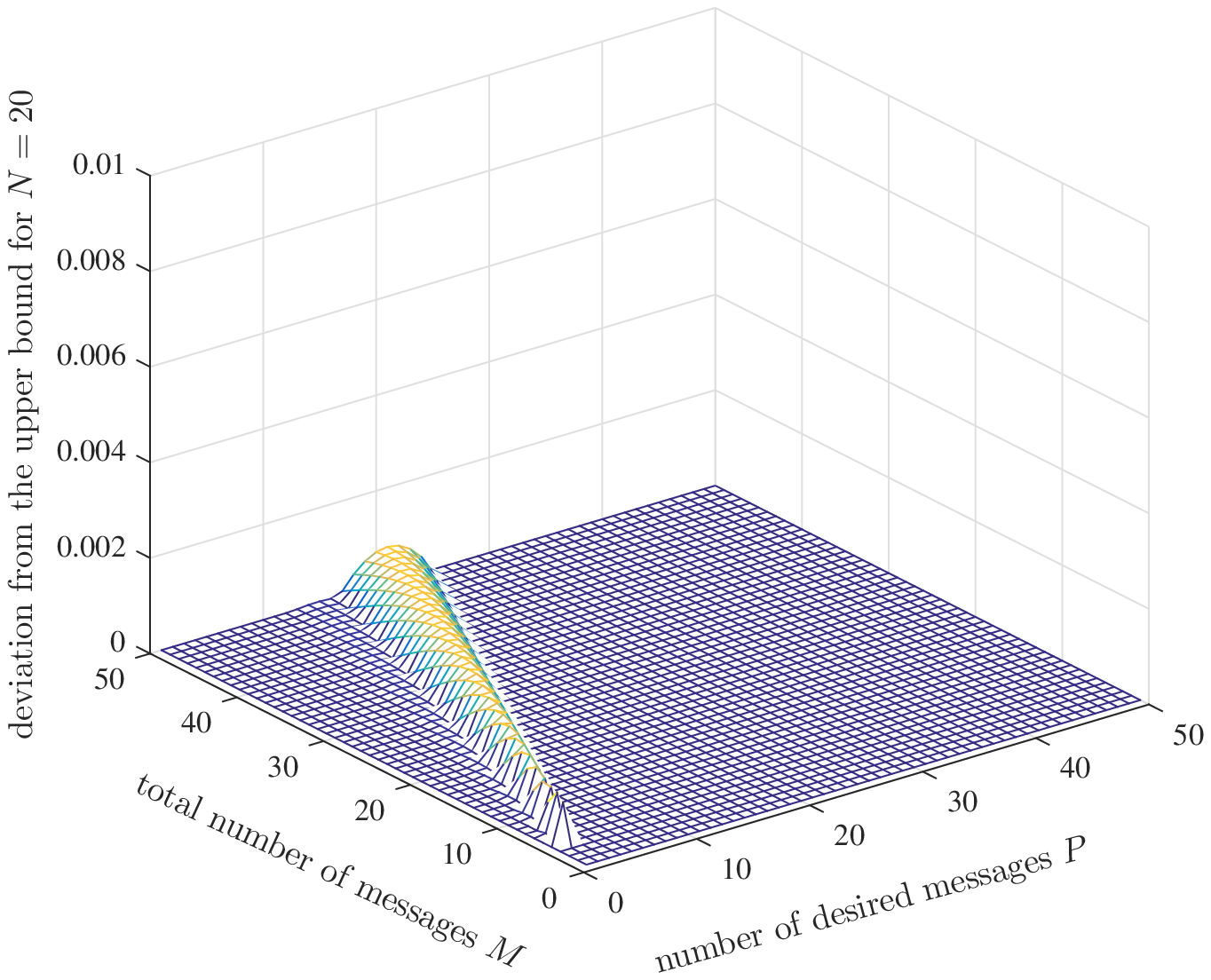}
\caption{$N=20$}
\end{subfigure}
\caption{Deviation of the achievable sum rate from the upper bound.}
\label{fig:3D_error}
\end{figure}

Fig.~\ref{fig:3D_error} shows the difference of the achievable rate $\underaccent{\bar}{R}_s$ and the upper bound $\bar{R}_s$ in Theorem~\ref{thm2}. The figure shows that the difference decreases as $N$ increases. This difference in all cases is small and is upper bounded by $0.0082$, which occurs when $M=5$, $P=2$, $N=2$. In addition, the difference is zero for the cases $P \geq \frac{M}{2}$ (Theorem~\ref{thm1}) or $\frac{M}{P} \in \mathbb{N}$ (Corollary~\ref{corollary3}).

\begin{figure}[t]
\centering
\includegraphics[width=0.6\textwidth]{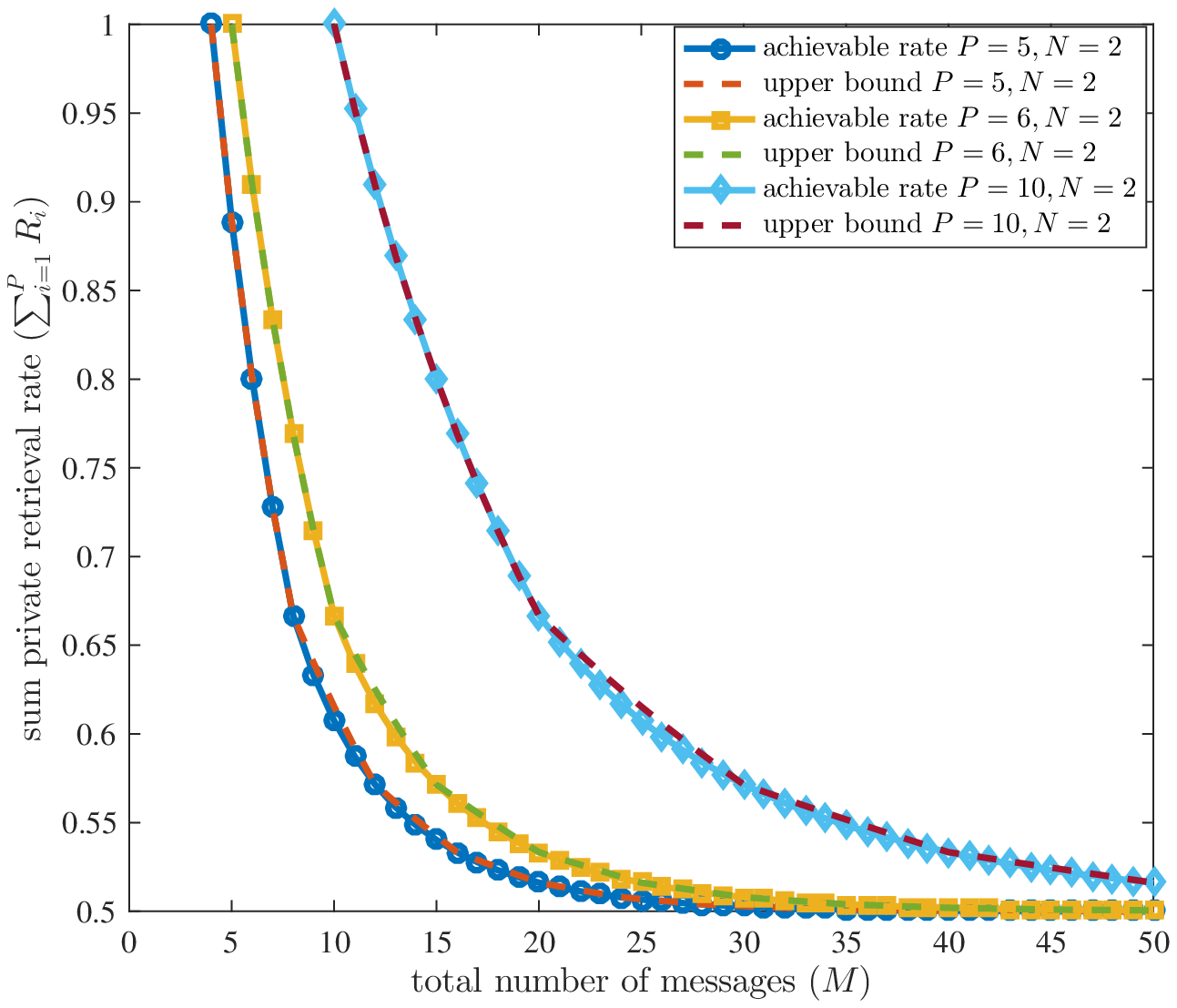}
\caption{Effect of changing $M$ for fixed $P=5,6,10$ and fixed $N=2$.}
\label{fig:M_slice}
\end{figure}

\begin{figure}[t]
\centering
\includegraphics[width=0.6\textwidth]{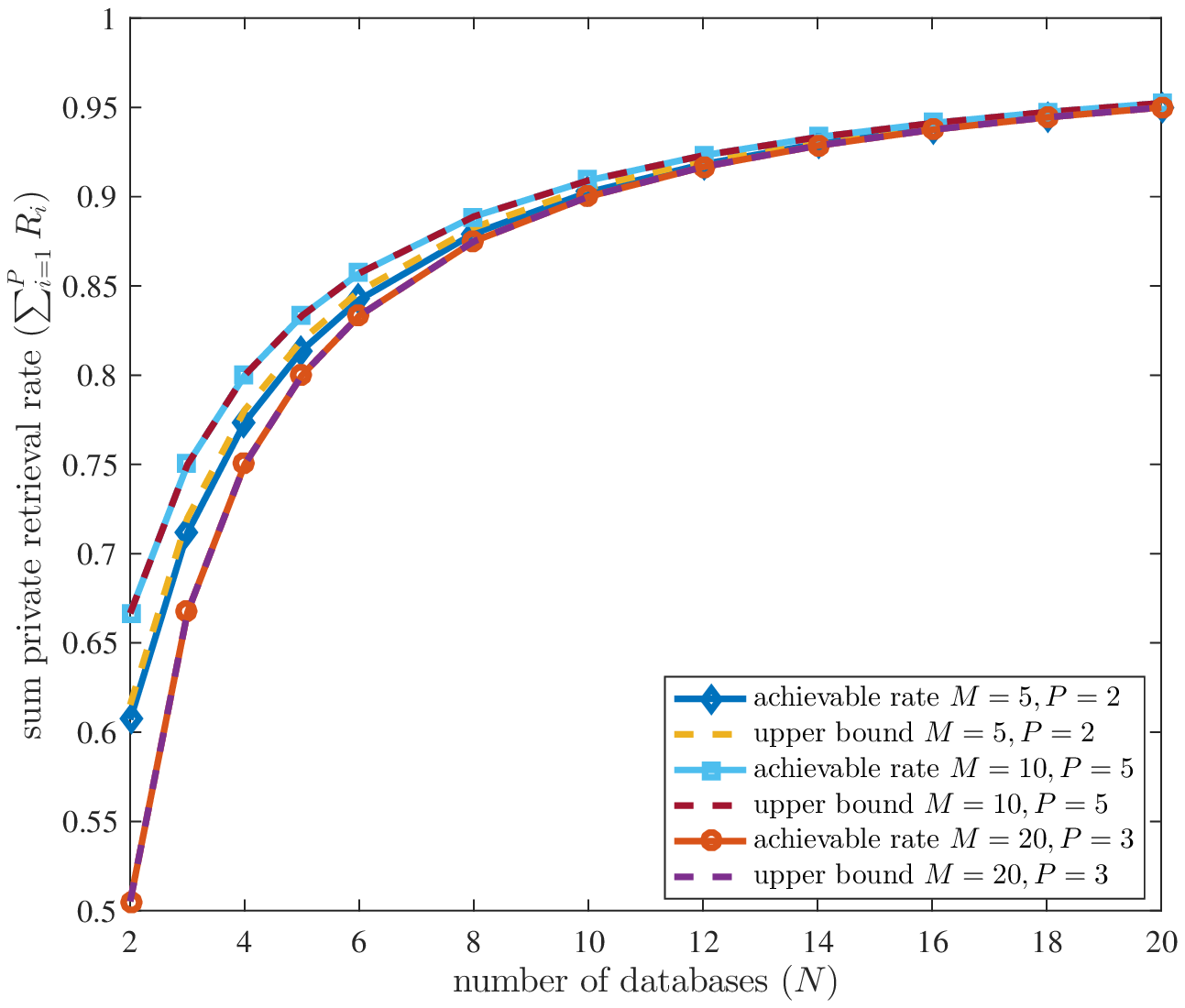}
\caption{Effect of changing $N$ for fixed $(M,P)=(5,2),(10,5),(20,3)$.}
\label{fig:N_slice}
\end{figure}

Fig.~\ref{fig:M_slice} shows the effect of changing $M$ for fixed $(P,N)$. We observe that as $M$ increases, the sum rate monotonically decreases and has a limit of $1-\frac{1}{N}$. In addition, Fig.~\ref{fig:N_slice} shows the effect of changing $N$ for fixed $(P,M)$. We observe that as $N$ increases, the sum rate increases and approaches $1$, as expected.

\section{Achievability Proof for the Case $P \geq \frac{M}{2}$} \label{sec:achievable1}

In this section, we present the general achievable scheme that attains the upper bound for the case $P \geq \frac{M}{2}$. The scheme applies the concepts of message symmetry, database symmetry, and exploiting side information as in \cite{JafarPIR}. However, our scheme requires the extra ingredient of MDS coding of the desired symbols and the side information in its second stage.

\subsection{Motivating Example: $M=3$, $P=2$ Messages, $N=2$ Databases} \label{mot1}

We start with a simple motivating example in this sub-section. The scheme operates over message size $N^2=4$. For sake of clarity, we assume that the three messages after interleaving their indices are $W_1=(a_1, \cdots, a_4)^T$, $W_2=(b_1, \cdots, b_4)^T$, and $W_3=(c_1, \cdots, c_4)^T$. We use $\bg_{2 \times 3}$ Reed-Solomon generator matrix over $\mathbb{F}_3$ as
\begin{align}
\bg_{2 \times 3}=\begin{bmatrix}
1 & 1 & 1 \\
1 & 2 & 3
\end{bmatrix}
\end{align}
The user picks a random permutation for the columns of $\bg_{2 \times 3}$ from the 6 possible permutations, e.g., in this example we use the permutation $2,1,3$. In the first round, the user starts by downloading one symbol from each database and each message, i.e., the user downloads $(a_1,b_1,c_1)$ from the first database, and $(a_2,b_2,c_2)$ from the second database. In the second round, the user encodes the side information from database 2 which is $c_2$ with two new symbols from $W_1,W_2$ which are $(a_3,b_3)$ using the permuted generator matrix, i.e., the user downloads two equations from database 1 in the second round,
\begin{align}
\bg\bs_1\begin{bmatrix}
a_3\\b_3\\c_2
\end{bmatrix}=
\begin{bmatrix}
1 & 1 & 1 \\
1 & 2 & 3
\end{bmatrix}
\begin{bmatrix}
0 & 1 & 0 \\
1 & 0 & 0 \\
0 & 0 & 1
\end{bmatrix}
\begin{bmatrix}
a_3\\b_3\\c_2
\end{bmatrix}
=\begin{bmatrix}
a_3+b_3+c_2 \\
2a_3+b_3+3c_2
\end{bmatrix}
\end{align}
The user repeats this operation for the second database with $(a_4,b_4)$ as desired symbols and $c_1$ as the side information from the first database.

For the decodability: The user subtracts out $c_2$ from round two in the first database, then the user can decode $(a_3,b_3)$ from $a_3+b_3$ and $2a_3+b_3$. Similarly, by subtracting out $c_1$ from round two in the second database, the user can decode $(a_4,b_4)$ from $a_4+b_4$ and $2a_4+b_4$.

For the privacy: Single bit retrievals of $(a_1, b_1, c_1)$ and $(a_2, b_2, c_2)$ from the two databases in the first round satisfy message symmetry and database symmetry, and do not leak any information. In addition, due to the private shuffling of bit indices, the different coefficients of 1, 2 and 3 in front of the bits in the MDS-coded summations in the second round do not leak any information either; see a formal proof in Section~\ref{dec-pri}. To see the privacy constraint intuitively from another angle, we note that the user can alter the queries for the second database when the queries for the first database are fixed, when the user wishes to retrieve another set of two messages. For instance, if the user wishes to retrieve $(W_1,W_3)$ instead of $(W_1,W_2)$, it can alter the queries for the second database by changing every $c_2$ in the queries of the second database with $c_3$, $c_1$ with $c_4$, $b_2$ with $b_3$, and $b_4$ with $b_1$.

The query table for this case is shown in Table~\ref{table(3,2,2)} below. The scheme retrieves $a_1, \cdots, a_4$ and $b_1,\cdots, b_4$, i.e., 8 bits in 10 downloads (5 from each database). Thus, the achievable sum rate for this scheme is $\frac{8}{10}=\frac{4}{5}=\frac{1}{1+\frac{M-P}{PN}}$. If we use the single-message optimal scheme in \cite{JafarPIR}, which is given in \cite[Example~4.3]{JafarPIR} for this specific case, twice in a row to retrieve two messages, we achieve a sum rate of $\frac{20}{28}=\frac{5}{7} < \frac{4}{5}$ as discussed in the introduction.

\begin{table}[h]
	\centering
	\caption{The query table for the case $M=3,P=2,N=2$.}
	\label{table(3,2,2)}
	\begin{tabular}{|c|c|}
		\hline
		Database 1 & Database 2 \\
		\hline
		$a_1,b_1,c_1$ & $a_2,b_2,c_2$ \\
		\hline
		$a_3+b_3+c_2$ & $a_4+b_4+c_1$ \\
		$2a_3+b_3+3c_2$& $2a_4+b_4+3c_1$\\
		\hline
	\end{tabular}
\end{table}

\subsection{General Achievable Scheme}

The scheme requires $L=N^2$, and is completed in two rounds. The main ingredient of the scheme is MDS coding of the desired symbols and side information in the second round. The details of the scheme are as follows.

\begin{enumerate}
\item \textit{Index preparation:} The user interleaves the contents of each message randomly and independently from the remaining messages using a random interleaver $\pi_m(.)$ which is known privately to the user only, i.e.,
	\begin{align}
	x_m(i)=w_m(\pi_m(i)), \quad i \in \{1, \cdots, L\}
	\end{align}
	where $X_m=[x_m(1), \cdots, x_m(L)]^T$ is the interleaved message. Thus, the downloaded symbol $x_m(i)$ at any database appears to be chosen at random and independent from the desired message subset $\cp$.
	
	\item \textit{Round one:} As in \cite{JafarPIR}, the user downloads one symbol from every message from every database, i.e., the user downloads $(x_{1}(n), x_{2}(n), \cdots, x_{M}(n))$ from the $n$th database. This implements \textit{message symmetry}, \textit{symmetry across databases} and satisfies the privacy constraint.
	
	\item \textit{Round two:} The user downloads a coded mixture of new symbols from the desired messages and the undesired symbols downloaded from the other databases. Specifically,
	
    \begin{enumerate}
		\item The user picks an MDS generator matrix $\bg \in \mathbb{F}_q^{P \times M}$, which has the property that every $P \times P$ submatrix is full-rank. This implies that if the user can cancel out any $M-P$ symbols from the mixture, the remaining symbols can be decoded. One explicit MDS generator matrix is the Reed-Solomon generator matrix over $\mathbb{F}_q$, where $q>M$, \cite{Reed_Solomon,wicker1999reed}
		\begin{align}
		\bg=\begin{bmatrix}
		1 & 1 & 1 & \cdots & 1 \\
		1 & 2 & 3 & \cdots & M \\
		1^2&2^2&3^2& \cdots & M^2 \\
		\vdots&\vdots&\vdots& \vdots&\vdots \\
		1^{P-1}&2^{P-1}&3^{P-1}& \cdots &M^{P-1}
		\end{bmatrix}_{P \times M}
		\end{align}
		
		\item The user picks uniformly and independently at random the permutation matrices $\bs_1, \bs_2, \cdots, \bs_{N-1}$ of size $M \times M$. These matrices shuffle the order of columns of $\bg$ to be independent of $\cp$.
		
		\item At the first database, the user downloads an MDS-coded version of $P$ new symbols from the desired set $\cp$ and $M-P$ undesired symbols that are already decoded from the second database in the first round, i.e., the user downloads $P$ equations of the form
		\begin{align}
		\bg\bs_1 	\begin{bmatrix}
		x_{i_1}(n+1) & x_{i_2}(n+1) & \cdots & x_{i_P} (n+1) & x_{j_1}(2) & x_{j_2}(2) & \cdots & x_{j_{M-P}}(2)
		\end{bmatrix}^T
		\end{align}
		where $\cp=\{i_1,i_2, \cdots, i_P\}$ are the indices of the desired messages and $\bar{\cp}=\{j_1,j_2, \cdots, j_{M-P}\}$ are the indices of the undesired messages. In this case, the user can cancel out the undesired messages and be left with a $P \times P$ invertible system of equations that it can solve to get $[x_{i_1}(n+1), x_{i_2}(n+1),\cdots,x_{i_P} (n+1)]$. This implements \textit{exploiting side information} as in \cite{JafarPIR}.
		
		\item The user repeats the last step for each set of side information from database 3 to database $N$, each with different permutation matrix.
		
		\item By \textit{database symmetry}, the user repeats all steps of round two at all other databases.
	\end{enumerate}
\end{enumerate}

\subsection{Decodability, Privacy, and Calculation of the Achievable Rate} \label{dec-pri}

Now, we verify that this achievable scheme satisfies the reliability and privacy constraints.

For the reliability: The user gets individual symbols from all databases in the first round, and hence they are all decodable by definition. In the second round, the user can subtract out all the undesired message symbols using the undesired symbols downloaded from all other databases during the first round. Consequently, the user is left with a $P \times P$ system of equations which is guaranteed to be invertible by the MDS property, hence all symbols that belong to $W_\cp$ are decodable.

For the privacy: At each database, for every message subset $\cp$ of size $P$, the achievable scheme retrieves randomly interleaved symbols which are encoded by the following matrix:
\begin{align}
\mathbf{H}_\cp=\begin{bmatrix}
\mathbf{I}_P & \mathbf{0}_P & \mathbf{0}_P & \cdots & \mathbf{0}_P \\
\mathbf{0}_P   & \bg^1_\cp  & \mathbf{0}_P & \cdots & \mathbf{0}_P \\
\mathbf{0}_P   & \mathbf{0}_P & \bg^2_\cp  & \cdots & \mathbf{0}_P \\
\vdots     &  \vdots    &   \vdots&   \vdots & \vdots \\
\mathbf{0}_P& \mathbf{0}_P& \mathbf{0}_P& \cdots& \bg^{N-1}_\cp
\end{bmatrix}
\end{align}
where $\bg^n_\cp=\bg\bs_n(:,\cp)$ are the columns of the encoding matrix that correspond to the message subset $\cp$ after applying the random permutation $\bs_n$. Since the permutation matrices are chosen uniformly and independently from each other, the probability distribution of $\mathbf{H}_\cp$ is uniform irrespective to $\cp$ (the probability of realizing such a matrix is $\left(\frac{(M-P)!}{M!}\right)^{N-1}$). Furthermore, the symbols are chosen randomly and uniformly by applying the random interleaver. Hence, the retrieval scheme is private.

To calculate the achievable rate: We note that at each database, the user downloads $M$ individual symbols in the first round that includes $P$ desired symbols. The user exploits the side information from the remaining $(N-1)$ databases to generate $P$ equations for each side information set. Each set of $P$ equations in turn generates $P$ desired symbols. Hence, the achievable rate is calculated as,
\begin{align}
\sum_{i=1}^P R_i&=\frac{\text{total number of desired symbols}}{\text{total downloaded equations}} \\
                &=\frac{N(P+P(N-1))}{N(M+P(N-1))} \\
                &=\frac{PN}{(M-P)+PN} \\
                &=\frac{1}{1+\frac{M-P}{PN}}
\end{align}

\subsection{Further Examples for the Case $P \geq \frac{M}{2}$}

In this section, we illustrate our achievable scheme with two more basic examples. In Section~\ref{mot1}, we considered the case $M=3$, $P=2$, $N=2$. In the next two sub-sections, we will consider examples with larger $M$, $P$ (Section~\ref{ex1}), and larger $N$ (Section~\ref{ex2}).

\subsubsection{$M=5$ Messages, $P=3$ Messages, $N=2$ Databases} \label{ex1}

Let $\cp=\{1,2,3\}$, and $a$ to $e$ denote the contents of $W_1$ to $W_5$, respectively. The achievable scheme is similar to the example in Section~\ref{mot1}. The difference is the use $5 \times 5$ permutation matrix for $\bs_1$ and $\bg_{3 \times 5}$ Reed-Solomon generator matrix over $\mathbb{F}_5$ as:
\begin{align}
\bg_{3 \times 5}=\begin{bmatrix}
1 & 1 & 1  & 1 & 1\\
1 & 2 & 3  & 4 & 5\\
1 & 4 & 4  & 1 & 0
\end{bmatrix}
\end{align}
The query table is shown in Table~\ref{table(5,3,2)} below with the following random permutation for the columns: $2,5,1,3,4$. The reliability and privacy constraints are satisfied due to the MDS property that implies that any subset of $3$ messages corresponds to a $3 \times 3$ invertible submatrix if the remaining symbols are decodable from the other database. This scheme retrieves $a_1,\cdots,a_4$, $b_1,\cdots,b_4$ and $c_1,\cdots,c_4$, hence 12 bits in 16 downloads (8 from each database). Thus, the achievable sum rate is $\frac{12}{16}=\frac{3}{4}$ which equals the sum capacity $\frac{1}{1+\frac{M-P}{PN}}$ in (\ref{result1}). This strictly outperforms the repetition-based achievable sum rate $\frac{18}{31}$ in (\ref{repeat}).

\begin{table}[h]
	\centering
	\caption{The query table for the case $M=5,P=3,N=2$.}
	\label{table(5,3,2)}
	\begin{tabular}{|c|c|}
		\hline
		Database 1 & Database 2 \\
		\hline
		$a_1,b_1,c_1,d_1,e_1$ & $a_2,b_2,c_2,d_2,e_2$ \\
		\hline
		$a_3+b_3+c_3+d_2+e_2$ & $a_4+b_4+c_4+d_1+e_1$ \\
		$2a_3+5b_3+c_3+3d_2+4e_2$& $2a_4+5b_4+c_4+3d_1+4e_1$\\
		$4a_3+c_3+4d_2+e_2$& $4a_4+c_4+4d_1+e_1$\\
		\hline
	\end{tabular}
\end{table}

\subsubsection{$M=4$ Messages, $P=2$ Messages, $N=3$ Databases}\label{ex2}

Next, we give an example with a larger $N$. Here, the message size is $N^2=9$. With a generator matrix $\bg_{2 \times 4}=\bg_{3 \times 5}([1:2],[1:4])$ to be the upper left submatrix of the previous example and two set of random permutations (corresponding to $\bs_1,\bs_2$) as $1,3,2,4$, and $4,1,3,2$. The query table is shown in Table~\ref{table(4,2,3)} below. This scheme retrieves $a_1,\cdots,a_9$ and $b_1,\cdots,b_9$, hence 18 bits in 24 downloads (8 from each database). Thus, the achievable rate is $\frac{18}{24}=\frac{3}{4}=\frac{1}{1+\frac{M-P}{PN}}$. This strictly outperforms the repetition-based achievable scheme sum rate $\frac{7}{10}$ in (\ref{repeat}).

\begin{table}
	\centering
	\caption{The query table for the case $M=4,P=2,N=3$.}
	\label{table(4,2,3)}
	\begin{tabular}{|c|c|c|}
		\hline
		Database 1 & Database 2 & Database 3\\
		\hline
		$a_1,b_1,c_1,d_1$ & $a_2,b_2,c_2,d_2$& $a_3,b_3,c_3,d_3$\\
		\hline
		$a_4+b_4+c_2+d_2$ & $a_6+b_6+c_1+d_1$&$a_8+b_8+c_1+d_1$ \\
		$a_4+3b_4+2c_2+4d_2$ & $a_6+3b_6+2c_1+4d_1$&$a_8+3b_8+2c_1+4d_1$ \\
		$a_5+b_5+c_3+d_3$ & $a_7+b_7+c_3+d_3$&$a_9+b_9+c_2+d_2$ \\
		$4a_5+b_5+3c_3+2d_3$ & $4a_7+b_7+3c_3+2d_3$&$4a_9+b_9+3c_2+2d_2$ \\
		\hline
	\end{tabular}
\end{table}

\section{Achievability Proof for the Case $P \leq \frac{M}{2}$}\label{sec:achievable2}

In this section, we describe an achievable scheme for the case $P \leq \frac{M}{2}$. We show that this scheme is optimal when the total number of messages $M$ is an integer multiple of the number of desired messages $P$. The scheme incurs a small loss from the upper bound for all other cases. The scheme generalizes the ideas in \cite{JafarPIR}. Different than \cite{JafarPIR}, our scheme uses unequal number of stages for each round of download. Interestingly, the number of stages at each round can be thought of as the output of an all-poles IIR filter. Our scheme reduces to \cite{JafarPIR} if we let $P=1$. In the sequel, we define the $i$th round as the download queries that retrieve sum of $i$ different symbols. We define the stage as a block of queries that exhausts all $\binom{M}{i}$ combinations of the sum of $i$ symbols in the $i$th round.

\subsection{Motivating Example: $M=5$, $P=2$ Messages, $N=2$ Databases} \label{mot2}

To motivate our achievable scheme, consider the case of retrieving two messages denoted by letters $(a,b)$ from five stored messages denoted by letters $(a, b, c, d, e)$. Instead of designing the queries beginning from the top as usual, i.e., beginning by downloading individual symbols, we design the scheme backwards starting from the last round that corresponds to downloading sums of all five messages and trace back to identify the  side information needed at each round from the other database. Our steps described below can be followed through in the query table in Table~\ref{table(5,2,2)}.

Now, let us fix the number of stages in the 5th round to be 1 as in \cite{JafarPIR} since $N=2$. Round 5 corresponds to downloading the sum of all five messages and contains one combination of symbols $a+b+c+d+e$; please see the last line in Table~\ref{table(5,2,2)}. Since we wish to retrieve $(a,b)$, we need one side information equation in the form of $c+d+e$ from earlier rounds. The combination $c+d+e$ can be created directly from round 3 without using round 4. Hence, we suppress round 4, as it does not create any useful side information in our case, and download one stage from round 3 to generate one side information equation $c+d+e$.

In round 3, we download sums of $3$ messages. Each stage of round 3 consists of $\binom{5}{3}=10$ equations. One of those 10 equations is in the desired $c+d+e$ form, and the remaining 9 of them have either $a$ or $b$ or both $a,b$ in them. In tabulating all these 9 combinations, we recognize two categories of side information equations needed from earlier rounds. The first category corresponds to equations of the form $a+b+(c,d,e)$, where $(c,d,e)$ means possible choices for the rest of the equation, i.e., these equations have both $a$ and $b$ in them and plus one more symbol in the form of $c$ or $d$ or $e$. This category requires downloading one stage of individual symbols (i.e., an individual $c$ or $d$ or $e$), that is, one stage of round 1. We note also that one of the symbols $(a,b)$ should be known as a side information from the second database in order to solve for the remaining new symbol. The second category corresponds to equations of the form $a+(c+d,c+e,d+e)$ and $b+(c+d,c+e,d+e)$, i.e., these equations have only one of $a$ or $b$ but not both. This category requires two stages of round 2, as we need different side information equations that contain sum of twos, e.g., $c+d$, $c+e$, $d+e$.

In round 2, we download sums of $2$ messages. Each stage of the second round contains $\binom{5}{2}=10$ equations. In each stage, we need one category of side information equations, which is $a+(c,d,e)$ and $b+(c,d,e)$. This necessitates two different stages of individual symbols, i.e., two stages of round 1 for each stage of round 2.

Denoting $\alpha_i$ to be the number of stages needed for the $i$th round, we sum all the required stages for round 1 to be $\alpha_1=2\cdot2+1=5$ stages. Hence, the user identifies the number stages as $\alpha_1=5, \alpha_2=2, \alpha_3=1, \alpha_4=0, \alpha_5=1$. These can be observed in the query table in Table~\ref{table(5,2,2)}. Note that, we have $\alpha_1=5$ stages in round 1 where we download individual bits; then we have $\alpha_2=2$ stages in round 2 where we download sums of two symbols; then we have $\alpha_3=1$ stage in round 3 where we download sums of three symbols; we skip round 4 as $\alpha_4=0$; and we have $\alpha_5=1$ stage of round 5 where we download sum of all five symbols.

Now, after designing the structure of the queries and the number of stages needed for each round, we apply the rest of the scheme described in \cite{JafarPIR}. The user randomly interleaves the messages as usual. In the first round, the user downloads one symbol from each message at each database. This is repeated $\alpha_1=5$ times for each database. Hence, the user downloads $a_{1:10}, b_{1:10},c_{1:10},d_{1:10},e_{1:10}$ from the two databases. In the second round, the user downloads sums of two messages. Each stage contains $\binom{5}{2}=10$ equations. This is repeated $\alpha_2=2$ times. For example, in database 1, user exploits $c_6,d_6,e_6$ to get $a_{12},a_{13},a_{14}$ and $c_7,d_7,e_7$ to obtain $b_{11},b_{12},b_{13}$. These are from round 1. Round 2 generates $c_{11}+d_{11}$, $c_{12}+e_{11}$, $d_{12}+e_{12}$ from stage 1, and $c_{13}+d_{13}$, $c_{14}+e_{13}$, $d_{14}+e_{14}$ from stage 2 as side information for round 3. In round 3, the user downloads sum of three symbols. There are $\binom{5}{3}=10$ of them. Symbols $c_{10},d_{10},e_{10}$ downloaded from round 1 in database 2 are used to be summed with mixtures of $a+b$. The two sets of side information generated in the second round are exploited in the equations that have one $a$ or $b$. Note that for each such equation, one of $a$ or $b$ is new and the other one is decoded from database 2. Round 3 generates one side information as $c_{19}+d_{19}+e_{19}$ that is used in round 5. This last round includes the sum of all five messages.

Therefore, as seen in Table~\ref{table(5,2,2)}, we have retrieved $a_1,\cdots,a_{34}$ and $b_1\cdots,b_{34}$, i.e., 68 bits in a total of 112 downloads (56 from each database). Thus, the achievable sum rate is $\frac{68}{112}=\frac{17}{28}$. This is $\underaccent{\bar}{R}_s$ in Theorem~\ref{thm2}, whereas the upper bound $\bar{R}_s$ in Theorem~\ref{thm2} is   $\frac{1}{1+\frac{1}{N}+\frac{1}{2N^2}}=\frac{8}{13}$. The gap between $\underaccent{\bar}{R}_s$ and $\bar{R}_s$ is equal to $\frac{3}{364}\simeq 0.0082$, which also is the largest possible gap between $\underaccent{\bar}{R}_s$ and $\bar{R}_s$ over all possible values of $M$, $P$ and $N$.

\begin{table}[]
	\centering
	\caption{The query table for the case $M=5,P=2,N=2$.}
	\label{table(5,2,2)}
	\begin{tabular}{|l|l|c|c|}
		\hline
		&                     & Database 1                          & Database 2                           \\ \hline
		\multirow{5}{*}{\rotatebox[origin=c]{90}{\parbox[c]{2cm}{\centering round 1}}}  & stg~1                        & $a_1,b_1,c_1,d_1,e_1$               & $a_6,b_6,c_6,d_6,e_6$                \\ \cline{2-4}
		& stg~2                        & $a_2,b_2,c_2,d_2,e_2$               & $a_7,b_7,c_7,d_7,e_7$                \\ \cline{2-4}
		& stg~3                        & $a_3,b_3,c_3,d_3,e_3$               & $a_8,b_8,c_8,d_8,e_8$                \\ \cline{2-4}
		& stg~4                        & $a_4,b_4,c_4,d_4,e_4$               & $a_9,b_9,c_9,d_9,e_9$                \\ \cline{2-4}
		& stg~5                        & $a_5,b_5,c_5,d_5,e_5$               & $a_{10},b_{10},c_{10},d_{10},e_{10}$ \\ \hline\hline
		\multirow{20}{*}{\rotatebox[origin=c]{90}{\parbox[c]{2cm}{\centering round 2}}} & \multirow{10}{*}{\rotatebox[origin=c]{90}{\parbox[c]{2cm}{\centering stage 1}}} & $a_{11}+b_6$                        & $a_{18}+b_1$                         \\
		&                           & $a_{12}+c_6$                        & $a_{19}+c_1$                         \\
		&                           & $a_{13}+d_6$                        & $a_{20}+d_1$                         \\
		&                           & $a_{14}+e_6$                        & $a_{21}+e_1$                         \\
		&                           & $b_{11}+c_7$                        & $b_{18}+c_2$                         \\
		&                           & $b_{12}+d_7$                        & $b_{19}+d_2$                         \\
		&                           & $b_{13}+e_7$                        & $b_{20}+e_2$                         \\
		&                           & $c_{11}+d_{11}$                     & $c_{15}+d_{15}$                      \\
		&                           & $c_{12}+e_{11}$                     & $c_{16}+e_{15}$                      \\
		&                           & $d_{12}+e_{12}$                     & $d_{16}+e_{16}$                      \\ \cline{2-4}
		& \multirow{10}{*}{\rotatebox[origin=c]{90}{\parbox[c]{2cm}{\centering stage 2}}} & $a_{6}+b_{14}$                      & $a_{1}+b_{21}$                       \\
		&                           & $a_{15}+c_8$                        & $a_{22}+c_3$                         \\
		&                           & $a_{16}+d_8$                        & $a_{23}+d_3$                         \\
		&                           & $a_{17}+e_8$                        & $a_{24}+e_3$                         \\
		&                           & $b_{15}+c_9$                        & $b_{22}+c_4$                         \\
		&                           & $b_{16}+d_9$                        & $b_{23}+d_4$                         \\
		&                           & $b_{17}+e_9$                        & $b_{24}+e_4$                         \\
		&                           & $c_{13}+d_{13}$                     & $c_{17}+d_{17}$                      \\
		&                           & $c_{14}+e_{13}$                     & $c_{18}+e_{17}$                      \\
		&                           & $d_{14}+e_{14}$                     & $d_{18}+e_{18}$                      \\ \hline\hline
		\multirow{10}{*}{\rotatebox[origin=c]{90}{\parbox[c]{2cm}{\centering round 3}}} & \multirow{10}{*}{\rotatebox[origin=c]{90}{\parbox[c]{2cm}{\centering stage 1}}} & $a_{25}+b_{7}+c_{10}$               & $a_{2}+b_{29}+c_{5}$                 \\
		&                           & $a_{7}+b_{25}+d_{10}$               & $a_{30}+b_{2}+d_{5}$                 \\
		&                           & $a_{26}+b_{8}+e_{10}$               & $a_{3}+b_{30}+e_{5}$                 \\
		&                           & $a_{27}+c_{15}+d_{15}$              & $a_{31}+c_{11}+d_{11}$               \\
		&                           & $a_{28}+c_{16}+e_{15}$              & $a_{32}+c_{12}+e_{11}$               \\
		&                           & $a_{29}+d_{16}+e_{16}$              & $a_{33}+d_{12}+e_{12}$               \\
		&                           & $b_{26}+c_{17}+d_{17}$              & $b_{31}+c_{13}+d_{13}$               \\
		&                           & $b_{27}+c_{18}+e_{17}$              & $b_{32}+c_{14}+e_{13}$               \\
		&                           & $b_{28}+d_{18}+e_{18}$              & $b_{33}+d_{14}+e_{14}$               \\
		&                           & $c_{19}+d_{19}+e_{19}$              & $c_{20}+d_{20}+e_{20}$               \\ \hline\hline
		\rotatebox[origin=c]{90}{\parbox[c]{1cm}{\centering rd.~5}}                 & stg~1                        & $a_{8}+b_{34}+c_{20}+d_{20}+e_{20}$ & $a_{34}+b_{3}+c_{19}+d_{19}+e_{19}$  \\ \hline
	\end{tabular}
\end{table}

\subsection{Calculation of the Number of Stages}\label{sec:stages}

The main new ingredient of our scheme in comparison to the scheme in \cite{JafarPIR} is the unequal number of stages in each round. In \cite{JafarPIR}, the scheme is completed in $M$ rounds, and each round contains only 1 stage only when $N=2$. To generalize the ideas in Section~\ref{mot2} and calculate the number of stages needed per round, we use Vandermonde's identity
\begin{align}
\binom{M}{i}=\sum_{k=0}^{P}\binom{P}{k}\binom{M-P}{i-k} \label{vander}
\end{align}
The relation in (\ref{vander}) states that any combination of $i$ objects from a group of $M$ objects must have $k$ objects from a group of size $P$ and $i-k$ objects from a group of size $M-P$. In our context, the first group is the subset of the desired messages and the second group is the subset of the undesired messages. Then, the relation can be interpreted in our setting as follows: In the $i$th round, the $\binom{M}{i}$ combinations of all possible sums of $i$ terms can be sorted into $P+1$ categories: The first category (i.e., $k=0$), contains no terms from the desired messages, the second category contains 1 term from the desired messages and $i-1$ terms from the undesired messages, etc. The relation gives also the number of query subgroups of each category $\binom{P}{k}$ and the number of queries in each subgroup $\binom{M-P}{i-k}$.

\textit{Let us consider the following concrete example for clarification:} Consider that we have $6$ messages denoted by $(a,b,c,d,e,f)$, and the desired group to be retrieved is $(a,b)$. Consider round 4 that consists of all combinations of sums of 4 symbols. From Vandermonde's identity, we know that $\binom{6}{4}=\binom{2}{0}\binom{4}{4} + \binom{2}{1}\binom{4}{3} + \binom{2}{2}\binom{4}{2}$. Which means that there are three categories of sums: First category is with only undesired messages; we have $\binom{2}{0}=1$ query subgroup of the form $c+d+e+f$. The second category is to have 1 term from the desired group and the remaining are undesired; we have $\binom{2}{1}=2$ query subgroups, one corresponds to $a$ with combinations of 3 terms from $c,d,e,f$, and the other to $b$ with combinations of 3 terms from $c,d,e,f$. Each query subgroup contains $\binom{4}{3}$ queries, i.e., the first query subgroup is of the form $a+(c+d+e, c+d+f, c+e+f , d+e+f)$ and the second query subgroup is of the form $b+(c+d+e, c+d+f, c+e+f , d+e+f)$. Third category is to have 2 terms from the desired group and 2 terms from the undesired group; we have $\binom{2}{2}=1$ query subgroup of this category that takes the form $a+b+(c+d, c+e, \cdots)$. The number of queries of this group is $\binom{4}{2}$ corresponding to all combinations of 2 undesired symbols.

\textit{Back to the calculation of the number of stages:} To be able to cancel the undesired symbols from an $i$-term sum, the user needs to download these undesired symbols as side information in the previous rounds. Hence, round $i$ requires downloading $\binom{P}{1}$ stages in round $(i-1)$, $\binom{P}{2}$ stages in round $(i-2)$, etc. Note that these stages need to be downloaded from the remaining $(N-1)$ databases. Then, each database needs to download $\frac{1}{N-1}\binom{P}{1}$ stages in round $(i-1)$, $\frac{1}{N-1}\binom{P}{2}$ stages in round $(i-2)$, etc.

From this observation, we can trace back the number of stages needed at each round. Denote $\alpha_i$ to be the number of stages in round $i$. Fix the number of stages in the last round (round $M$) to be $\alpha_M=(N-1)^{M-P}$ stages. This choice ensures that the number of stages in any round is an integer. Note that in round $M$, the user downloads a sum of all $M$ messages, this requires side information in the form of the sum of the undesired $M-P$ messages. Hence, we suppress the rounds $M-P+1$ through $M-1$ since they do not generate any useful side information. Note that the side information equations in round $M$ at each database are collected from the remaining $(N-1)$ databases. Then, the number of stages in round $(M-P)$ should be $(N-1)^{M-P-1}$. Therefore, we write
\begin{align}
\alpha_M&=(N-1)^{M-P} \\
\alpha_{M-1}&=\cdots= \alpha_{M-P+1}=0 \\
\alpha_{M-P}&=(N-1)^{M-P-1}=\frac{1}{N-1} \alpha_M=\frac{1}{N-1}\sum_{i=1}^{P} \binom{P}{i} \alpha_{M-P+i}
\end{align}

Now, in round $(M-P)$, each stage requires $\binom{P}{1}$ stages from round $(M-P-1)$, $\binom{P}{2}$ stages from round $(M-P-2)$, and so on so forth, and these stages are divided across $(N-1)$ databases. Continuing with the same argument, for each round, we write
\begin{align}
\alpha_{M-P-1}&=\frac{1}{N-1}\binom{P}{1}\alpha_{M-P}=\frac{1}{N-1}\sum_{i=1}^{P} \binom{P}{i} \alpha_{M-P-1+i} \\
\alpha_{M-P-2}&=\frac{1}{N-1}\binom{P}{1}\alpha_{M-P-1}+\frac{1}{N-1}\binom{P}{2}\alpha_{M-P}=\frac{1}{N-1}\sum_{i=1}^{P} \binom{P}{i} \alpha_{M-P-2+i}\\
\vdots \nonumber \\
\alpha_k&=\frac{1}{N-1}\sum_{i=1}^{P} \binom{P}{i} \alpha_{k+i}
\end{align}
Interestingly, this pattern closely resembles the output of an IIR filter $y[n]$ \cite{oppenheim}, with the difference equation,
\begin{align}\label{IIR}
y[n]=\frac{1}{N-1}\sum_{i=1}^{P} \binom{P}{i} y[n-i]
\end{align}
and with the initial conditions $y[-P]=(N-1)^{M-P},\: y[-P+1]= \cdots=y[-1]=0$. Note that the only difference between the two seemingly different settings is the orientation of the time axis. The calculation of the number of stages is obtained backwards in contrast to the output of this IIR filter. Hence, we can systematically obtain the number of stages at each round by observing the output of the IIR filter characterized by (\ref{IIR}), and mapping it to the number of stages via $\alpha_k=y[(M-P)-k]$.

We note that for the special case $P=1$, the number of stages can be obtained from the first order filter $y[n]=\frac{1}{N-1}y[n-1]$. The output of this filter is $y[n]=(N-1)^{M-2-n}$. Then, the number of stages in round $k$ is $\alpha_k=y[M-1-k]=(N-1)^{k-1}$, which is exactly the number of stages used in \cite{JafarPIR}; in particular if $N=2$, then $\alpha_k=1$ for all $k$.

\subsection{General Achievable Scheme}

\begin{enumerate}
	\item \textit{Index preparation:} The user interleaves the contents of each message randomly and independently from the remaining messages using a random interleaver $\pi_m(.)$ which is known privately to the user only, i.e.,
	\begin{align}
	x_m(i)=w_m(\pi_m(i)), \quad i \in \{1, \cdots, L\}
	\end{align}
	
	\item \textit{Number of stages:} We calculate the number of stages needed in each round. This can be done systematically by finding the output of the IIR filter characterized by,
	\begin{align}\label{filter}
	y[n]=\frac{1}{N-1}\sum_{i=1}^{P} \binom{P}{i} y[n-i]
	\end{align}
	with the initial conditions $y[-P]=(N-1)^{M-P},\: y[-P+1]= \cdots=y[-1]=0$. The number of stages in round $i$ is $\alpha_i=y[(M-P)-i]$ as discussed in Section~\ref{sec:stages}.
	
	\item \textit{Initialization:} From the first database, the user downloads one symbol from each message that belongs to the desired message set $\cp$. The user sets the round index to $i=1$.
	
	\item \textit{Message symmetry:} In round $i$, the user downloads sums of $i$ terms from different symbols from the first database. To satisfy the privacy constraint, the user should download an equal amount of symbols from all messages. Therefore, the user downloads the remaining $\binom{M-P}{i}$ combinations in round $i$ from the undesired symbol set $\bar{\cp}$. For example: In round 1, the user downloads one symbol from every undesired message with a total of $\binom{M-P}{1}=M-P$ such symbols.
	
	\item \textit{Repetition of stages:} In the first database, the user repeats the operation in round $i$ according to the number of calculated stages $\alpha_i$. This in total results in downloading $\alpha_i\binom{M-P}{i}$ undesired equations, and $\alpha_i \left(\binom{M}{i}-\binom{M-P}{i}\right)$ desired equations.
	
	\item \textit{Symmetry across databases:} The user implements symmetry across databases by downloading $\alpha_i\binom{M-P}{i}$ new undesired equations, and $\alpha_i\left(\binom{M}{i}-\binom{M-P}{i}\right)$ new desired equations from each database. These undesired equations will be used as side information in subsequent rounds. For example: In round 1, each database generates $\alpha_1(M-P)$ undesired equations in the form of individual symbols. Hence, each database can exploit up to $\alpha_1 (N-1)(M-P)$ side information equations from other $(N-1)$ databases.
	
	\item \textit{Exploiting side information:} Until now, we did not specify how the desired equations are constructed. Since each stage in round $i$ can be categorized using Vandermonde's identity as in the previous section, we form the desired equations as a sum of the desired symbols and the undesired symbols that can be decoded from other databases in the former $(i-1)$ rounds. If the user sums two or more symbols from $\cp$, the user downloads one new symbol from one message only and the remaining symbols from $\cp$ should be derived from other databases. Thus, in round $(i+1)$, the user mixes one symbol of $\cp$ with the sum of $i$ undesired symbols from round $i$. This should be repeated for all $\binom{P}{1}$ desired symbols. Then, the user mixes each sum of 2 desired symbols with the sum of $(i-1)$  undesired symbols generated in the $(i-1)$th round. This should be repeated for all the $\binom{P}{2}$ combinations of the desired symbols, and so on.
	
	\item \textit{Repeating steps:} Repeat steps 4, 5, 6, 7 by setting $i=i+1$ until $i=M-P-1$.
	
	\item \textit{Last round:} We note that rounds $M-P+1$ to $M-1$ do not generate useful side information. Hence, $\alpha_{M-P+1}= \cdots=\alpha_{M-1}=0$. In round $M$, which corresponds to summing all $M$ messages, the user mixes $P$ symbols from $\cp$ (only one of them is new and the remaining are previously decoded from the other $(N-1)$ databases) and $M-P$ undesired symbol mixture that was generated in round $(M-P)$.
	
	\item \textit{Shuffling the order of queries:} After preparing the query table, the order of the queries are shuffled uniformly, so that all possible orders of queries are equally likely regardless of $\cp$.
\end{enumerate}

\subsection{Decodability, Privacy, and Calculation of the Achievable Rate}

Now, we verify that the proposed scheme satisfies the reliability and privacy constraints.

For the reliability: The scheme is designed to download the \emph{exact} number of undesired equations that will be used as side information equation at subsequent rounds in other databases.\footnote{Check for instance in Table~\ref{table(5,2,2)} that all of the downloads (equations) involving undesired symbols from database 2 are used in database 1: singles $c_6, d_6, e_6$, $c_7, d_7, e_7$, $c_8, d_8, e_8$, $c_9, d_9, e_9$, $c_{10}, d_{10}, e_{10}$; sums of twos $c_{15}+d_{15}$, $c_{16}+e_{15}$, $d_{16}+e_{16}$, $c_{17}+d_{17}$, $c_{18}+e_{17}$, $d_{18}+e_{18}$; sum of threes $c_{20}+d_{20}+e_{20}$, all downloaded from database 2 are all used as side information in database 1.} Hence, each desired symbol at any round is mixed with a known mixture of symbols that can be decoded from other databases. Note that if the scheme encounters the case of having a mixture of desired symbols, one of them only is chosen to be new and the remaining symbols are downloaded previously from other databases. Thus, the reliability constraint is satisfied by canceling out the side information.

For the privacy: The randomized mapping of message bits and the randomization of the order of queries guarantees privacy as in \cite{JafarPIR}. It can be checked that when we fix the queries for one database, we can adjust the queries for the remaining databases such that the user can decode any $\cp$ subset of messages. This is true since all combinations of messages are generated by our scheme.

To calculate the achievable rate: From Vandermonde's identity $\binom{M}{i}=\sum_{p=0}^P \binom{P}{p}\binom{M-P}{i-p}$, round $i$ requires downloading $\binom{P}{p}$ stages in round $(i-p)$. These stages should be downloaded from the remaining $(N-1)$ databases. Hence, as shown in the previous section, the number of stages at each round is calculated as the output of an IIR filter whose input-output relation is given in (\ref{IIR}) with the initial conditions $y[-P]=(N-1)^{M-P},\: y[-P+1]=\cdots=y[-1]=0$, with the conversion of time index of the filter to the round index of the schemes as $\alpha_i=y[(M-P)-i]$. These initial conditions imply that the user downloads $(N-1)^{M-P}$ stages in the last round that corresponds to downloading the sum of all messages. The $(P-1)$ rounds before the last round are suppressed because we only need to form sums of $(M-P)$ messages to be used in the last round.

Now, to calculate the number of stages for round $i$, we first solve for the roots of the characteristic equation of (\ref{IIR}) \cite{oppenheim},
\begin{align}
r^P-\frac{1}{N-1}\sum_{i=1}^P \binom{P}{i} r^{P-i}=0
\end{align}
which is equivalent to
\begin{align}
r^P-\frac{r^P}{N-1}\sum_{i=1}^P \binom{P}{i} r^{-i}=0
\end{align}
which further reduces to
\begin{align} \label{ccs_binom}
r^P-\frac{r^P}{N-1}\left[\left(1+\frac{1}{r}\right)^P-1\right]=0
\end{align}
using the binomial theorem. Simplifying (\ref{ccs_binom}), we have
\begin{align} \label{ccs_simplified}
Nr^P-(r+1)^P=0
\end{align}
By applying the bijective mapping $t=N^{1/P} \cdot\frac{r}{r+1}$, (\ref{ccs_simplified}) is equivalent to $t^P=1$. The roots for this equation are the normal roots of unity, i.e., $t_k=e^{j2\pi(k-1)/P}, \: k=1, \cdots, P$, where $j=\sqrt{-1}$. Hence, the roots of the characteristic equation are given by,
\begin{align}
r_k=\frac{t_k}{N^{1/P}-t_k}=\frac{e^{j2\pi(k-1)/P}}{N^{1/P}-e^{j2\pi(k-1)/P}}, \quad  k=1, \cdots, P
\end{align}
Thus, the complete response of the IIR filter is given by $y[n]=\sum_{i=1}^{P} \gamma_i r_i^n$, where $\gamma_i$ are constants that result from solving the initial conditions, i.e., $\boldsymbol{\gamma}=(\gamma_1, \cdots, \gamma_P)^T$ is the solution of the system of equations,
\begin{align}
\begin{bmatrix}
r_1^{-P} & r_2^{-P} & \cdots & r_P^{-P} \\
r_1^{-P+1} & r_2^{-P+1} & \cdots & r_P^{-P+1} \\
\vdots   &  \vdots  &  \cdots& \vdots \\
r_1^{-1} & r_2^{-1} & \cdots & r_P^{-1} \\
\end{bmatrix}
\begin{bmatrix}
\gamma_1 \\\gamma_2 \\ \vdots\\ \gamma_P
\end{bmatrix}
=
\begin{bmatrix}
(N-1)^{M-P} \\ 0 \\ \vdots \\ 0
\end{bmatrix}
\end{align}

Now, we are ready to calculate the number of stages $\alpha_k$ in round $k$. Since $\alpha_k=y[(M-P)-k]$ by construction, then
\begin{align}
\alpha_k=\sum_{i=1}^{P} \gamma_i r_i^{M-P-k}
\end{align}
In round $k$, the user downloads sums of $k$ symbols. The user repeats this round for $\alpha_k$ stages. Each stage contains all the combinations of any $k$ symbols which there are $\binom{M}{k}$ of them. Hence, the total download cost $D$ is,
\begin{align}
D&=\sum_{k=1}^M \binom{M}{k} \alpha_k \\
 &=\sum_{k=1}^M \sum_{i=1}^P \binom{M}{k} \gamma_i r_i^{M-P-k}\\
 &=\sum_{i=1}^P \gamma_i r_i^{M-P} \sum_{k=1}^{M} \binom{M}{k} r_i^{-k}\\
 &=\sum_{i=1}^P \gamma_i r_i^{M-P}\left[\left(1+\frac{1}{r_i}\right)^M-1\right]
\end{align}
Considering the undesired equations: in round $k$, the user downloads all combinations of the $(M-P)$ undesired messages which there are $\binom{M-P}{k}$ of them. Therefore, similar to the above calculation, the total number of undesired equations $U$ is,
\begin{align}
U=\sum_{i=1}^P \gamma_i r_i^{M-P}\left[\left(1+\frac{1}{r_i}\right)^{M-P}-1\right]
\end{align}
Hence, the achievable rate $\underaccent{\bar}{R}_s$ is
\begin{align}\label{rate1}
\underaccent{\bar}{R}_s &=\frac{D-U}{D} \\
             &=\frac{\sum_{i=1}^{P}\gamma_i r_i^{M-P}\left[\left(1+\frac{1}{r_i}\right)^M-\left(1+\frac{1}{r_i}\right)^{M-P}\right]}{\sum_{i=1}^{P}\gamma_i r_i^{M-P}\left[\left(1+\frac{1}{r_i}\right)^M-1\right]}
\end{align}
which is (\ref{ach-rate}) in Theorem~\ref{thm2}.

\subsection{Further Examples for the Case $P \leq \frac{M}{2}$}

In this section, we illustrate our proposed scheme with a few additional basic examples. In Section~\ref{mot2}, we considered the case $M=5$, $P=2$, $N=2$. In the next three sub-sections, we consider three more examples. In the example in Section~\ref{ex3}, the ratio $\frac{M}{P}$ is exactly equal to 2, thus, both the achievable scheme here and the achievable scheme in Section~\ref{sec:achievable1} can be used; we comment about the differences and advantages of both schemes. In the example in Section~\ref{ex4}, we present the case of a larger $N$ for the example in Section~\ref{mot2}. In the example in Section~\ref{ex5}, we present a case with larger $M$, $P$ and $N$.

\subsubsection{$M=4$ Messages, $P=2$ Messages, $N=2$ Databases} \label{ex3}

The first step of the achievable scheme is to identify the number of stages needed for each round of download. The IIR filter in (\ref{IIR}) that determines the number of stages reduces in this case to
\begin{align}
y[n]=2y[n-1]+y[n-2]
\end{align}
with the initial conditions $y[-2]=1,y[-1]=0$. The number of stages in round $k$ is $\alpha_k=y[2-k]$. Since $M$ is small, we can calculate the output iteratively without using the canonical filter output as,
\begin{align}
\alpha_4&=y[-2]=1 \\
\alpha_3&=y[-1]=0 \\
\alpha_2&=y[0]=2y[-1]+y[-2]=1 \\
\alpha_1&=y[1]=2y[0]+y[-1]=2
\end{align}
Hence, we should download 2 stages of individual symbols (round 1), and 1 stage of sums of two symbols (round 2). We should suppress the round that retrieves sums of three symbols (round 3), and have 1 stage of sums of all four symbols (round 4).

The user initializes the scheme by randomly and independently interleaving the symbols of each message. The query table for this example is shown in Table~\ref{table(4,2,2)}. In round 1, the user downloads individual symbols from all messages at each database. The user downloads $a_1,b_1,c_1,d_1$ and $a_2,b_2,c_2,d_2$ from database 1, as $\alpha_1=2$. This is repeated for database 2. In round 2, the user downloads sums of two symbols. There are $\binom{4}{2}=6$ such equations. At database 1, the undesired symbols from database 2 in the first round are exploited in some of these sums. These equations are either in the form $a+(c,d)$ or in the form $b+(c,d)$. This necessitates two sets of different individual symbols to be downloaded from database 2 in the first round, or otherwise the symbols are repeated and privacy is compromised. Moreover, we note that the user downloads $a_5+b_3$ which uses $b_3$ as side information even though $W_2$ is desired; this is reversed in database 2 to download $a_1+b_7$ with $a_1$ as a side information to have a symmetric scheme. Round 2 concludes with downloading $c_5+d_5$ and $c_6+d_6$ at the two databases, which will be used as side information in the last round. Round 3 is skipped and the user proceeds to round 4 (last round) directly. In round 4, the user downloads sum of four symbols, and uses the side information downloaded in round 2 and any decoded symbols for the other desired message. For example, in database 1, the user downloads $a_3+b_{10}+c_6+d_6$, hence, the side information $c_6+d_6$ is exploited in this round as well as $a_3$. The user finishes the scheme by shuffling the order of all queries randomly. The user retrieves $a_1,\cdots,a_{10}$ and $b_1,\cdots,b_{10}$ privately in 30 downloads (15 from each database) and achieves a sum rate of $\frac{20}{30}=\frac{2}{3}=\frac{1}{1+\frac{1}{N}}$, which matches the upper bound in Theorem~\ref{thm2}. This sum rate outperforms the repetition-based achievable rate which is $\frac{3}{5}$ in (\ref{repeat}).

\begin{table}[t]
	\centering
	\caption{The query table for the case $M=4,P=2,N=2$.}
	\label{table(4,2,2)}
	\begin{tabular}{|l|l|c|c|}
		\hline
		&                          & Database 1             & Database 2             \\ \hline
		\multirow{2}{*}{\rotatebox[origin=c]{90}{\parbox[c]{0.9cm}{\centering rd.~1}}} & stg~1                     & $a_1,b_1,c_1,d_1$      & $a_3,b_3,c_3,d_3$      \\ \cline{2-4}
		& stg~2                     & $a_2,b_2,c_2,d_2$      & $a_4,b_4,c_4,d_4$      \\ \hline\hline
		\multirow{6}{*}{\rotatebox[origin=c]{90}{\parbox[c]{2cm}{\centering round 2}}} & \multirow{6}{*}{\rotatebox[origin=c]{90}{\parbox[c]{2cm}{\centering stage 1}}} & $a_5+b_3$              & $a_1+b_7$              \\
		&                          & $a_6+c_3$              & $a_8+c_1$              \\
		&                          & $a_7+d_3$              & $a_9+d_1$              \\
		&                          & $b_5+c_4$              & $b_8+c_2$              \\
		&                          & $b_6+d_4$              & $b_9+d_2$              \\
		&                          & $c_5+d_5$              & $c_6+d_6$              \\ \hline\hline
		\rotatebox[origin=c]{90}{\parbox[c]{1cm}{\centering rd.~4}}                    & stg~1                     & $a_{3}+b_{10}+c_6+d_6$ & $a_{10}+b_{1}+c_5+d_5$ \\ \hline
	\end{tabular}
\end{table}

We note that this case can be solved using the achievable scheme presented in Section~\ref{sec:achievable1} as well since $\frac{M}{P}=2$ in this case. In fact, this is equivalent to the case considered in Section~\ref{ex2}, if the number of databases is reduced from $N=3$ to $N=2$. Starting from Table~\ref{table(4,2,3)} in Section~\ref{ex2} and removing the downloads from database 3, we obtain the query table which uses MDS-coded queries shown in Table~\ref{table(4,2,2)alt} below. Via the scheme in Table~\ref{table(4,2,2)alt} below, the user retrieves $a_1,\cdots,a_4$ and $b_1,\cdots,b_4$ privately in 12 downloads (6 from each database), therefore achieving the same optimal sum rate of $\frac{8}{12}=\frac{2}{3}=\frac{1}{1+\frac{1}{N}}$.

\begin{table}[h]
	\centering
	\caption{Alternative query table for the case $M=4,P=2,N=2$.}
	\label{table(4,2,2)alt}
	\begin{tabular}{|c|c|}
		\hline
		Database 1 & Database 2 \\
		\hline
		$a_1,b_1,c_1,d_1$ & $a_2,b_2,c_2,d_2$ \\
		\hline
		$a_3+b_3+c_2+d_2$ & $a_4+b_4+c_1+d_1$ \\
		$a_3+3b_3+2c_2+4d_2$ & $a_4+3b_4+2c_1+4d_1$ \\
		\hline
	\end{tabular}
\end{table}

We presented this case here even though it could be solved using the scheme in Section~\ref{sec:achievable1}, in order to give an example where the second achievable scheme achieves the upper bound in Theorem~\ref{thm2} and yields a capacity result since $\frac{M}{P}$ is an integer. Interestingly, we observe that for all cases where $P=\frac{M}{2}$, the two  achievable schemes are both optimal. The two schemes present an interesting trade-off between the field size and the upload cost: The first achievable scheme in Section~\ref{sec:achievable1} requires using an MDS code with field size $q \geq M$ but the number of queries for each database is limited to $M+P$. On the other hand, the second achievable scheme here in Section~\ref{sec:achievable2} does not use any coding and can work with the storage field size, however, the number of queries increase exponentially since the number of stages for each round is related to an unstable IIR filter.

\subsubsection{$M=5$ Messages, $P=2$ Messages, $N=3$ Databases} \label{ex4}

In this example, we show an explicit query structure for $N>2$. In this case the corresponding difference equation for the IIR filter is
\begin{align}
y[n]=y[n-1]+\frac{1}{2}y[n-2]
\end{align}
with the initial conditions $y[-1]=0,\,y[-2]=(N-1)^{M-P}=8$. Thus, the number of stages in each round are: $\alpha_1=6$, $\alpha_2=4$, $\alpha_3=4$, $\alpha_4=0, \alpha_5=8$. The query table is shown in Tables~\ref{table(5,2,3)}, \ref{table(5,2,3)_cont} and \ref{table(5,2,3)_cont2}. This scheme retrieves $a_1,\cdots,a_{126}$ and $b_1,\cdots,b_{126}$ privately in 354 downloads (177 from each database), therefore, achieving a sum rate of $\frac{252}{354}=\frac{42}{59}<\frac{1}{1+\frac{1}{N}+\frac{1}{2N^2}}=\frac{18}{25}$. The gap is $\frac{12}{1475} \simeq 0.0081$.

\begin{table}[]
	\centering
	\caption{The query table for the case $M=5,P=2,N=3$.}
	\label{table(5,2,3)}
	\begin{tabular}{|l|l|c|c|c|}
		\hline
		&                           & Database 1                      & Database 2                           & Database 3                           \\ \hline
		
		\multirow{6}{*}{\rotatebox[origin=c]{90}{\parbox[c]{2cm}{\centering round 1}}}  & stg~1                      & $a_1,b_1,c_1,d_1,e_1$           & $a_7,b_7,c_7,d_7,e_7$                & $a_{13},b_{13},c_{13},d_{13},e_{13}$ \\ \cline{2-5}
		& stg~2                      & $a_2,b_2,c_2,d_2,e_2$           & $a_8,b_8,c_8,d_8,e_8$                & $a_{14},b_{14},c_{14},d_{14},e_{14}$ \\ \cline{2-5}
		& stg~3                      & $a_3,b_3,c_3,d_3,e_3$           & $a_9,b_9,c_9,d_9,e_9$                & $a_{15},b_{15},c_{15},d_{15},e_{15}$ \\ \cline{2-5}
		& stg~4                      & $a_4,b_4,c_4,d_4,e_4$           & $a_{10},b_{10},c_{10},d_{10},e_{10}$ & $a_{16},b_{16},c_{16},d_{16},e_{16}$ \\ \cline{2-5}
		& stg~5                      & $a_5,b_5,c_5,d_5,e_5$           & $a_{11},b_{11},c_{11},d_{11},e_{11}$ & $a_{17},b_{17},c_{17},d_{17},e_{17}$ \\ \cline{2-5}
		& stg~6                      & $a_{6},b_{6},c_{6},d_{6},e_{6}$ & $a_{12},b_{12},c_{12},d_{12},e_{12}$ & $a_{18},b_{18},c_{18},d_{18},e_{18}$ \\ \hline\hline
		\multirow{30}{*}{\rotatebox[origin=c]{90}{\parbox[c]{2cm}{\centering round 2}}} & \multirow{10}{*}{\rotatebox[origin=c]{90}{\parbox[c]{2cm}{\centering stage 1}}} & $a_{19}+b_7$                    & $a_{33}+b_1$                         & $a_{47}+b_{1}$                       \\
		&                           & $a_{20}+c_7$                    & $a_{34}+c_1$                         & $a_{48}+c_1$                         \\
		&                           & $a_{21}+d_7$                    & $a_{35}+d_1$                         & $a_{49}+d_1$                         \\
		&                           & $a_{22}+e_7$                    & $a_{36}+e_1$                         & $a_{50}+e_1$                         \\
		&                           & $b_{19}+c_8$                    & $b_{33}+c_2$                         & $b_{47}+c_2$                         \\
		&                           & $b_{20}+d_8$                    & $b_{34}+d_2$                         & $b_{48}+d_2$                         \\
		&                           & $b_{21}+e_8$                    & $b_{35}+e_2$                         & $b_{49}+e_2$                         \\
		&                           & $c_{19}+d_{19}$                 & $c_{27}+d_{27}$                      & $c_{35}+d_{35}$                      \\
		&                           & $c_{20}+e_{19}$                 & $c_{28}+e_{27}$                      & $c_{36}+e_{35}$                      \\
		&                           & $d_{20}+e_{20}$                 & $d_{28}+e_{28}$                      & $d_{36}+e_{36}$                      \\ \cline{2-5}
		& \multirow{10}{*}{\rotatebox[origin=c]{90}{\parbox[c]{2cm}{\centering stage 2}}} & $a_{7}+b_{22}$                  & $a_{1}+b_{36}$                       & $a_{1}+b_{50}$                       \\
		&                           & $a_{23}+c_9$                    & $a_{37}+c_3$                         & $a_{51}+c_3$                         \\
		&                           & $a_{24}+d_9$                    & $a_{38}+d_3$                         & $a_{52}+d_3$                         \\
		&                           & $a_{25}+e_9$                    & $a_{39}+e_3$                         & $a_{53}+e_3$                         \\
		&                           & $b_{23}+c_{10}$                 & $b_{37}+c_4$                         & $b_{51}+c_4$                         \\
		&                           & $b_{24}+d_{10}$                 & $b_{38}+d_4$                         & $b_{52}+d_4$                         \\
		&                           & $b_{25}+e_{10}$                 & $b_{39}+e_4$                         & $b_{53}+e_4$                         \\
		&                           & $c_{21}+d_{21}$                 & $c_{29}+d_{29}$                      & $c_{37}+d_{37}$                      \\
		&                           & $c_{22}+e_{21}$                 & $c_{30}+e_{29}$                      & $c_{38}+e_{37}$                      \\
		&                           & $d_{22}+e_{22}$                 & $d_{30}+e_{30}$                      & $d_{38}+e_{38}$                      \\ \cline{2-5}
		& \multirow{10}{*}{\rotatebox[origin=c]{90}{\parbox[c]{2cm}{\centering stage 3}}} & $a_{26}+b_{13}$                 & $a_{40}+b_{13}$                      & $a_{54}+b_{7}$                       \\
		&                           & $a_{27}+c_{13}$                 & $a_{41}+c_{13}$                      & $a_{55}+c_7$                         \\
		&                           & $a_{28}+d_{13}$                 & $a_{42}+d_{13}$                      & $a_{56}+d_7$                         \\
		&                           & $a_{29}+e_{13}$                 & $a_{43}+e_{13}$                      & $a_{57}+e_7$                         \\
		&                           & $b_{26}+c_{14}$                 & $b_{40}+c_{14}$                      & $b_{54}+c_8$                         \\
		&                           & $b_{27}+d_{14}$                 & $b_{41}+d_{14}$                      & $b_{55}+d_8$                         \\
		&                           & $b_{28}+e_{14}$                 & $b_{42}+e_{14}$                      & $b_{56}+e_8$                         \\
		&                           & $c_{23}+d_{23}$                 & $c_{31}+d_{31}$                      & $c_{39}+d_{39}$                      \\
		&                           & $c_{24}+e_{23}$                 & $c_{32}+e_{31}$                      & $c_{40}+e_{39}$                      \\
		&                           & $d_{24}+e_{24}$                 & $d_{32}+e_{32}$                      & $d_{40}+e_{40}$                      \\ \cline{1-5}
	\end{tabular}
\end{table}

\begin{table}[]
	\centering
	\caption{The query table for the case $M=5,P=2,N=3$ (cont.).}
	\label{table(5,2,3)_cont}
	\begin{tabular}{|l|l|c|c|c|}
		\hline
		&                           & Database 1                            & Database 2                            & Database 3                            \\ \hline
		\multirow{10}{*}{\rotatebox[origin=c]{90}{\parbox[c]{2cm}{\centering round 2}}}		& \multirow{10}{*}{\rotatebox[origin=c]{90}{\parbox[c]{2cm}{\centering stage 4}}} & $a_{13}+b_{29}$                 & $a_{13}+b_{43}$                      & $a_{7}+b_{57}$                       \\
		&                           & $a_{30}+c_{15}$                 & $a_{44}+c_{15}$                      & $a_{58}+c_9$                         \\
		&                           & $a_{31}+d_{15}$                 & $a_{45}+d_{15}$                      & $a_{59}+d_9$                         \\
		&                           & $a_{32}+e_{15}$                 & $a_{46}+e_{15}$                      & $a_{60}+e_9$                         \\
		&                           & $b_{30}+c_{16}$                 & $b_{44}+c_{16}$                      & $b_{58}+c_{10}$                      \\
		&                           & $b_{31}+d_{16}$                 & $b_{45}+d_{16}$                      & $b_{59}+d_{10}$                      \\
		&                           & $b_{32}+e_{16}$                 & $b_{46}+e_{16}$                      & $b_{60}+e_{10}$                      \\
		&                           & $c_{25}+d_{25}$                 & $c_{33}+d_{33}$                      & $c_{41}+d_{41}$                      \\
		&                           & $c_{26}+e_{25}$                 & $c_{34}+e_{33}$                      & $c_{42}+e_{41}$                      \\
		&                           & $d_{26}+e_{26}$                 & $d_{34}+e_{34}$                      & $d_{42}+e_{42}$ \\ \hline\hline
		\multirow{30}{*}{\rotatebox[origin=c]{90}{\parbox[c]{2cm}{\centering round 3}}} & \multirow{10}{*}{\rotatebox[origin=c]{90}{\parbox[c]{2cm}{\centering stage 1}}} & $a_{61}+b_{8}+c_{11}$                 & $a_{79}+b_{2}+c_{5}$                  & $a_{97}+b_{2}+c_{5}$                  \\
		&                           & $a_{8}+b_{61}+d_{11}$                 & $a_{2}+b_{79}+d_{5}$                  & $a_{2}+b_{97}+d_{5}$                  \\
		&                           & $a_{62}+b_{9}+e_{11}$                 & $a_{80}+b_{3}+e_{5}$                  & $a_{98}+b_{3}+e_{5}$                  \\
		&                           & $a_{63}+c_{27}+d_{27}$                & $a_{81}+c_{19}+d_{19}$                & $a_{99}+c_{19}+d_{19}$                \\
		&                           & $a_{64}+c_{28}+e_{27}$                & $a_{82}+c_{20}+e_{19}$                & $a_{100}+c_{20}+e_{19}$               \\
		&                           & $a_{65}+d_{28}+e_{28}$                & $a_{83}+d_{20}+e_{20}$                & $a_{101}+d_{20}+e_{20}$               \\
		&                           & $b_{62}+c_{29}+d_{29}$                & $b_{80}+c_{21}+d_{21}$                & $b_{98}+c_{21}+d_{21}$                \\
		&                           & $b_{63}+c_{30}+e_{29}$                & $b_{81}+c_{22}+e_{21}$                & $b_{99}+c_{22}+e_{21}$                \\
		&                           & $b_{64}+d_{30}+e_{30}$                & $b_{82}+d_{22}+e_{22}$                & $b_{100}+d_{22}+e_{22}$               \\
		&                           & $c_{43}+d_{43}+e_{43}$                & $c_{47}+d_{47}+e_{47}$                & $c_{51}+d_{51}+e_{51}$                \\ \cline{2-5}
		& \multirow{10}{*}{\rotatebox[origin=c]{90}{\parbox[c]{2cm}{\centering stage 2}}} & $a_{9}+b_{65}+c_{12}$                 & $a_{3}+b_{83}+c_{6}$                  & $a_{3}+b_{101}+c_{6}$                 \\
		&                           & $a_{66}+b_{10}+d_{12}$                & $a_{84}+b_{4}+d_{6}$                  & $a_{102}+b_{4}+d_{6}$                 \\
		&                           & $a_{10}+b_{66}+e_{12}$                & $a_{4}+b_{84}+e_{6}$                  & $a_{4}+b_{102}+e_{6}$                 \\
		&                           & $a_{67}+c_{31}+d_{31}$                & $a_{85}+c_{23}+d_{23}$                & $a_{103}+c_{23}+d_{23}$               \\
		&                           & $a_{68}+c_{32}+e_{31}$                & $a_{86}+c_{24}+e_{23}$                & $a_{104}+c_{24}+e_{23}$               \\
		&                           & $a_{69}+d_{32}+e_{32}$                & $a_{87}+d_{24}+e_{24}$                & $a_{105}+d_{24}+e_{24}$               \\
		&                           & $b_{67}+c_{33}+d_{33}$                & $b_{85}+c_{25}+d_{25}$                & $b_{103}+c_{25}+d_{25}$               \\
		&                           & $b_{68}+c_{34}+e_{33}$                & $b_{86}+c_{26}+e_{25}$                & $b_{104}+c_{26}+e_{25}$               \\
		&                           & $b_{69}+d_{34}+e_{34}$                & $b_{87}+d_{26}+e_{26}$                & $b_{105}+d_{26}+e_{26}$               \\
		&                           & $c_{44}+d_{44}+e_{44}$                & $c_{48}+d_{48}+e_{48}$                & $c_{52}+d_{52}+e_{52}$                \\ \cline{2-5}
		& \multirow{10}{*}{\rotatebox[origin=c]{90}{\parbox[c]{2cm}{\centering stage 3}}} & $a_{70}+b_{14}+c_{17}$                & $a_{88}+b_{14}+c_{17}$                & $a_{106}+b_{8}+c_{11}$                \\
		&                           & $a_{14}+b_{70}+d_{17}$                & $a_{14}+b_{88}+d_{17}$                & $a_{8}+b_{106}+d_{11}$                \\
		&                           & $a_{71}+b_{15}+e_{17}$                & $a_{89}+b_{15}+e_{17}$                & $a_{107}+b_{9}+e_{11}$                \\
		&                           & $a_{72}+c_{35}+d_{35}$                & $a_{90}+c_{35}+d_{35}$                & $a_{108}+c_{27}+d_{27}$               \\
		&                           & $a_{73}+c_{36}+e_{35}$                & $a_{91}+c_{36}+e_{35}$                & $a_{109}+c_{28}+e_{27}$               \\
		&                           & $a_{74}+d_{36}+e_{36}$                & $a_{92}+d_{36}+e_{36}$                & $a_{110}+d_{28}+e_{28}$               \\
		&                           & $b_{71}+c_{37}+d_{37}$                & $b_{89}+c_{37}+d_{37}$                & $b_{107}+c_{29}+d_{29}$               \\
		&                           & $b_{72}+c_{38}+e_{37}$                & $b_{90}+c_{38}+e_{37}$                & $b_{108}+c_{30}+e_{29}$               \\
		&                           & $b_{73}+d_{38}+e_{38}$                & $b_{91}+d_{38}+e_{38}$                & $b_{109}+d_{30}+e_{30}$               \\
		&                           & $c_{45}+d_{45}+e_{45}$                & $c_{49}+d_{49}+e_{49}$                & $c_{53}+d_{53}+e_{53}$                \\ \cline{1-5}
	\end{tabular}
\end{table}

\begin{table}[]
	\centering
	\caption{The query table for the case $M=5,P=2,N=3$ (cont.).}
	\label{table(5,2,3)_cont2}
	\begin{tabular}{|l|l|c|c|c|}
		\hline
		&                           & Database 1                            & Database 2                            & Database 3                            \\ \hline
			\multirow{10}{*}{\rotatebox[origin=c]{90}{\parbox[c]{2cm}{\centering round 3}}}		
			& \multirow{10}{*}{\rotatebox[origin=c]{90}{\parbox[c]{2cm}{\centering stage 4}}} & $a_{15}+b_{74}+c_{18}$                & $a_{15}+b_{92}+c_{18}$                & $a_{9}+b_{110}+c_{12}$                \\
			&                           & $a_{75}+b_{16}+d_{18}$                & $a_{93}+b_{16}+d_{18}$                & $a_{111}+b_{10}+d_{12}$               \\
			&                           & $a_{16}+b_{75}+e_{18}$                & $a_{16}+b_{93}+e_{18}$                & $a_{10}+b_{111}+e_{12}$               \\
			&                           & $a_{76}+c_{39}+d_{39}$                & $a_{94}+c_{39}+d_{39}$                & $a_{112}+c_{31}+d_{31}$               \\
			&                           & $a_{77}+c_{40}+e_{39}$                & $a_{95}+c_{40}+e_{39}$                & $a_{113}+c_{32}+e_{31}$               \\
			&                           & $a_{78}+d_{40}+e_{40}$                & $a_{96}+d_{40}+e_{40}$                & $a_{114}+d_{32}+e_{32}$               \\
			&                           & $b_{76}+c_{41}+d_{41}$                & $b_{94}+c_{41}+d_{41}$                & $b_{112}+c_{33}+d_{33}$               \\
			&                           & $b_{77}+c_{42}+e_{41}$                & $b_{95}+c_{42}+e_{41}$                & $b_{113}+c_{34}+e_{33}$               \\
			&                           & $b_{78}+d_{42}+e_{42}$                & $b_{96}+d_{42}+e_{42}$                & $b_{114}+d_{34}+e_{34}$               \\
			&                           & $c_{46}+d_{46}+e_{46}$                & $c_{50}+d_{50}+e_{50}$                & $c_{54}+d_{54}+e_{54}$                \\ \hline\hline
			\multirow{8}{*}{\rotatebox[origin=c]{90}{\parbox[c]{2cm}{\centering round 5}}}  & stg~1                      & $a_{115}+b_{11}+c_{47}+d_{47}+e_{47}$ & $a_{119}+b_{5}+c_{43}+d_{43}+e_{43}$  & $a_{123}+b_{5}+c_{43}+d_{43}+e_{43}$  \\ \cline{2-5}
			& stg~2                      & $a_{11}+b_{115}+c_{48}+d_{48}+e_{48}$ & $a_{5}+b_{119}+c_{44}+d_{44}+e_{44}$  & $a_{5}+b_{123}+c_{44}+d_{44}+e_{44}$  \\ \cline{2-5}
			& stg~3                      & $a_{116}+b_{12}+c_{49}+d_{49}+e_{49}$ & $a_{120}+b_{6}+c_{45}+d_{45}+e_{45}$  & $a_{124}+b_{6}+c_{45}+d_{45}+e_{45}$  \\ \cline{2-5}
			& stg~4                      & $a_{12}+b_{116}+c_{50}+d_{50}+e_{50}$ & $a_{6}+b_{120}+c_{46}+d_{46}+e_{46}$  & $a_{6}+b_{124}+c_{46}+d_{46}+e_{46}$  \\ \cline{2-5}
			& stg~5                      & $a_{117}+b_{17}+c_{51}+d_{51}+e_{51}$ & $a_{121}+b_{17}+c_{51}+d_{51}+e_{51}$ & $a_{125}+b_{11}+c_{47}+d_{47}+e_{47}$ \\ \cline{2-5}
			& stg~6                      & $a_{17}+b_{117}+c_{52}+d_{52}+e_{52}$ & $a_{17}+b_{121}+c_{52}+d_{52}+e_{52}$ & $a_{11}+b_{125}+c_{48}+d_{48}+e_{48}$ \\ \cline{2-5}
			& stg~7                      & $a_{118}+b_{18}+c_{53}+d_{53}+e_{53}$ & $a_{122}+b_{18}+c_{53}+d_{53}+e_{53}$ & $a_{126}+b_{12}+c_{49}+d_{49}+e_{49}$ \\ \cline{2-5}
			& stg~8                      & $a_{18}+b_{118}+c_{54}+d_{54}+e_{54}$ & $a_{18}+b_{122}+c_{54}+d_{54}+e_{54}$ & $a_{12}+b_{126}+c_{50}+d_{50}+e_{50}$ \\ \hline
	\end{tabular}
\end{table}

\subsubsection{$M=7$ Messages, $P=3$ Messages, $N=3$ Databases} \label{ex5}

Finally, in this section, we consider an example with $N=3$ databases and larger $M$ and $P$ than in previous examples, where we describe the structure and the calculation of the number of queries without specifying the explicit query table as it grows quite large. We first calculate the number of stages at each round. The corresponding IIR filter is
\begin{align}
y[n]=\frac{1}{2}(3y[n-1]+3y[n-2]+y[n-3])
\end{align}
with the initial conditions $y[-3]=(N-1)^{M-P}=16$, $y[-2]=0$, $y[-1]=0$. Hence, the number of stages for each round $\alpha_k=y[4-k]$,  $k=1,\cdots, 7$, are calculated iteratively as $\alpha_1=67$, $\alpha_2=30$, $\alpha_3=12$, $\alpha_4=8$, $\alpha_5=0$, $\alpha_6=0$, $\alpha_7=16$.

In round 1, the user downloads $67$ individual symbols from each message and from each database. Each database can use the side information generated by the other two databases. Hence, each database has $67\cdot2=134$ side information equations in the form of single symbols from round 1 to exploit. In round 2, the user downloads sums of two symbols. Each stage in round 2 requires $3$ stages from round 1, since the user faces with $a+(d,e,f,g)$, $b+(d,e,f,g)$ or $c+(d,e,f,g)$ cases. Then, round 2 requires $30\cdot3=90$ stages from the generated side information in round 1, and we are left with $134-90=44$ more stages of round 1. Each database can use the side information stages from the other two databases, i.e., each can use up to $2\cdot30=60$ stages of side information in the form of sums of two.

In round 3, the user downloads sums of three symbols, which can be either of $a+b+(d,e,f,g)$, $a+c+(d,e,f,g)$, $b+c+(d,e,f,g)$, $a+(d+e,d+f,\cdots)$, and similarly for $b,c$. Therefore, each stage in round 3 requires $3$ stages from round 2, and $3$ stages from round 1. This in total requires $12\cdot3=36$ stages from round 1 and $36$ stages from round 2, and we will be left with $8$ stages from round 1 and $24$ stages from round 2. Round 3 generates $2\cdot12=24$ stages of side information in the form of sums of threes. In round 4, the user downloads sums of 4 symbols, which can be either $a+b+(d+e, d+f, \cdots)$, and similarly for $b+c$ and $a+c$, $a+(d+e+f,d+e+g, \cdots)$ and similarly for $b$, $c$, or $a+b+c+(d,e,f,g)$. This means that for each stage of round 3, the user needs 1 stage of round 1, 3 stages of round 2, and 3 stages of round 3. This in total requires $8\cdot3=24$ stages from round 2 and 3 and $8\cdot1$ stages from round 1 and hence, we exhaust all the generated side information by round 4. Round 4 generates $8$ stages of side information in the form of sums of fours. This will be used in the last round to get $8\cdot2$ new symbols from the desired messages.

The achievable sum rate in this case is $\frac{3933}{5445} =\frac{437}{605}< \frac{1}{1+\frac{1}{N}+\frac{1}{3N^2}} =\frac{27}{37}$. The gap is $\frac{166}{22385} \simeq 0.0074$.

\section{Converse Proof}

In this section, we derive an upper bound for the MPIR problem. The derived upper bound is tight when $P \geq \frac{M}{2}$ and when $\frac{M}{P} \in \mathbb{N}$. We follow the notations and simplifications in \cite{JafarPIR, KarimCoded}, and we define
\begin{align}
\cq&\triangleq \left\{Q_n^{[\cp]}: \cp \subseteq \{1, \cdots, M\}, \quad |\cp|=P,\: n \in \{1, \cdots, N\}\right\} \\
A_{n_1:n_2}^{[\cp]}&\triangleq\left\{A_{n_1}^{[\cp]}, A_{n_1+1}^{[\cp]}, \cdots, A_{n_2}^{[\cp]} \right\}, \quad n_1 \leq n_2, \: n_1,n_2 \in \{1, \cdots, N\}
\end{align}
Without loss of generality, the following simplifications hold for the MPIR problem:
\begin{enumerate}
\item We can assume that the MPIR scheme is symmetric. Since for every asymmetric scheme, there exists an equal rate symmetric scheme that can be constructed by replicating all permutations of databases and messages.
\item To invoke the privacy constraint, we fix the response of one database to be the same irrespective of the desired set of messages $\cp$, i.e., $A_n^{[\cp_i]}=A_n$, where $|\cp_i|=P$ for every $i \in \left\{1,2, \cdots, \beta\right\}$ for some $n \in \{1, \cdots, N\}$, and $\beta=\binom{M}{P}$. No loss of generality is incurred due to the fact that the queries and the answers are statistically independent from $\cp$. In the sequel, we fix the answer string of the first database, i.e.,
\begin{align}\label{fixed_answer}
A_1^{[\cp]}=A_1, \: \forall \cp
\end{align}
\end{enumerate}

The following lemma is a consequence of the symmetry assumption; its proof can be found in \cite{JafarPIR}.

\begin{lemma}[Symmetry \cite{JafarPIR}] \label{symmetry}
	For any $W_\mathcal{S}=\{W_i:i \in \mathcal{S}\}$
	\begin{align}
	H(A_n^{[\cp]}|W_\mathcal{S},\cq)&=H(A_1^{[\cp]}|W_\mathcal{S},\cq), \quad n \in\{1,\cdots, N\} \\
	H(A_1|\cq)&=H(A_n^{[\cp]}|\cq), \quad n \in \{1, \cdots, N\}, \: \forall \cp
	\end{align}
\end{lemma}

We construct the converse proof by induction over $\lfloor\frac{M}{P}\rfloor$ in a similar way to \cite{JafarPIR, KarimCoded}. The base induction step is obtained for $1 \leq \frac{M}{P} \leq 2$ (this is the case $P\geq\frac{M}{2}$ as it was referred to so far, where the user wants to retrieve at least half of the messages). We obtain an inductive relation for the case $\frac{M}{P}>2$. The converse proof extends the proof in \cite{JafarPIR} for $P >1$.

\subsection{Converse Proof for the Case $1 \leq \frac{M}{P} \leq 2$}\label{converse1}

To prove the converse for the case $1 \leq \frac{M}{P} \leq 2$, we need the following lemma which gives a lower bound on the interference within an answer string.

\begin{lemma}[Interference Lower Bound]\label{interference}
For the MPIR problem with $P \geq \frac{M}{2}$, the uncertainty of the interfering messages $W_{P+1:M}$ within the answer string $A_1^{[1:P]}$ is lower bounded as,
\begin{align}\label{lower_bound}
H(A_1^{[1:P]}|W_{1:P},\cq) \geq \frac{(M-P)L}{N}
\end{align}
Furthermore, (\ref{lower_bound}) is true for any set of desired messages $\cp$ with $|\cp|=P$, i.e.,
\begin{align}
H(A_1^{[\cp]}|W_{\cp},\cq) \geq \frac{(M-P)L}{N}
\end{align}
\end{lemma}

\begin{Proof}
For clarity of presentation, we assume that $\cp=\{1, \cdots, P\}$ without loss of generality. Hence,
	\begin{align}
  	    (M-P)L&=H(W_{P+1:M}) \\
\label{lb1}   &=H(W_{P+1:M}|W_{1:P},\cq) \\
\label{lb2}   &=H(W_{P+1:M}|W_{1:P},\cq)-H(W_{P+1:M}|A_{1:N}^{[M-P+1:M]},W_{1:P},\cq)\\
              &=I(W_{P+1:M};A_{1:N}^{[M-P+1:M]}|W_{1:P},\cq)\\
\label{lb3}   &=H(A_{1:N}^{[M-P+1:M]}|W_{1:P},\cq) \\
              &\leq \sum_{n=1}^N H(A_{n}^{[M-P+1:M]}|W_{1:P},\cq) \\
\label{lb4}   &=NH(A_1|W_{1:P},\cq)
	\end{align}
where (\ref{lb1}) follows from the independence of the messages $W_{P+1:M}$ from the messages $W_{1:P}$ and the queries as in (\ref{msg_indep}) and (\ref{msg_query_indep}); (\ref{lb2}) follows from the reliability constraint (\ref{reliability}), since messages $W_{P+1:M}$ can be decoded correctly from the answer strings $A_{1:N}^{[M-P+1:M]}$ if $P \geq \frac{M}{2}$ as $\{P+1, \cdots, M\} \subseteq \{M-P+1, \cdots, M\}$ in this regime; (\ref{lb3}) follows from the fact that the answer strings are deterministic functions of all messages and queries $(\cq,W_{1:M})$; and (\ref{lb4}) follows from the independence bound and Lemma~\ref{symmetry}.

Consequently, $H(A_1|W_{1:P},\cq) \geq \frac{(M-P)L}{N}$. The proof of the general statement can be done replacing $W_{1:P}$ by $W_{\cp}$, $W_{P+1:M}$ by $W_{\bar{\cp}}$ which corresponds to the complement set of messages of $W_{\cp}$, and the answer strings $A_{1:N}^{[M-P+1:M]}$ by $A_{1:N}^{[\cp^*]}$, where $\bar{\cp}\subseteq\cp^*, \: |\cp^*|=P$.
\end{Proof}

Now, we are ready to prove the converse of the case $P \geq \frac{M}{2}$. We use a similar converse technique to the case of $M=2, P=1$ in \cite{JafarPIR},
\begin{align}
		  ML&=H(W_{1:M})\\
\label{c1}  &=H(W_{1:M}|\cq) \\
\label{c2}  &=H(W_{1:M}|\cq)-H(W_{1:M}|A_{1:N}^{[\cp_1]},A_{1:N}^{[\cp_2]}, \cdots, A_{1:N}^{[\cp_{\beta}]},\cq) \\
            &=I(W_{1:M};A_{1:N}^{[\cp_1]},A_{1:N}^{[\cp_2]}, \cdots, A_{1:N}^{[\cp_{\beta}]}|\cq) \\
\label{c3}  &=H(A_{1:N}^{[\cp_1]},A_{1:N}^{[\cp_2]}, \cdots, A_{1:N}^{[\cp_{\beta}]}|\cq) \\
\label{c4}  &=H(A_1,A_{2:N}^{[\cp_1]},A_{2:N}^{[\cp_2]}, \cdots, A_{2:N}^{[\cp_\beta]}|\cq)\\
            &=H(A_1,A_{2:N}^{[\cp_1]}|\cq)+H(A_{2:N}^{[\cp_2]}, \cdots, A_{2:N}^{[\cp_\beta]}|A_1,A_{2:N}^{[\cp_1]},\cq)\\
\label{c5}  &=H(A_1,A_{2:N}^{[\cp_1]}|\cq)+H(A_{2:N}^{[\cp_2]}, \cdots, A_{2:N}^{[\cp_\beta]}|A_1,A_{2:N}^{[\cp_1]},W_\cp,\cq)\\
\label{c6}  &\leq \sum_{n=1}^N H(A_n^{[\cp]}|\cq)+H(A_{2:N}^{[\cp_2]}, \cdots, A_{2:N}^{[\cp_\beta]}|A_1,W_\cp,\cq)\\
\label{c7}  &=\sum_{n=1}^N H(A_n^{[\cp]}|\cq)+H(A_{1:N}^{[\cp_2]}, \cdots, A_{1:N}^{[\cp_\beta]}|W_\cp,\cq)-H(A_1|W_\cp,\cq)
\end{align}
where (\ref{c1}) follows from the independence between the messages and the queries; (\ref{c2}) follows from the reliability constraint in (\ref{reliability}) with  $A_{1:N}^{[\cp_1]},A_{1:N}^{[\cp_2]}, \cdots, A_{1:N}^{[\cp_{\beta}]}$ representing all answer strings from all databases to any subset of messages $\cp \subseteq \{1, \cdots, M\}$; (\ref{c3}) follows from the fact that answer strings are deterministic functions of the messages and the queries; (\ref{c4}) follows from simplification (\ref{fixed_answer}) without loss of generality; (\ref{c5}) follows from the fact that the messages $W_\mathcal{P}=(W_{i_1}, W_{i_2}, \cdots, W_{i_P})$ can be reconstructed from $A_{1:N}^{[\cp]}$; and (\ref{c6}) is a consequence of the fact that conditioning does not increase entropy and Lemma~\ref{symmetry}.

Now, every message appears in $\binom{M-1}{P-1}$ different message subsets of size $P$, therefore the answer strings $(A_{1:N}^{[\cp_2]}, \cdots, A_{1:N}^{[\cp_\beta]})$ are sufficient to construct all messages $W_{1:M}$ irrespective of $\cp_1$. Therefore,
\begin{align}
H(A_{1:N}^{[\cp_2]}, \cdots, A_{1:N}^{[\cp_\beta]}|W_\cp,\cq)=(M-P)L
\end{align}
Using this and Lemma~\ref{interference} in (\ref{c7}) yields
\begin{align}
ML &\leq \sum_{n=1}^N H(A_n^{[\cp]}|\cq)+(M-P)L-\frac{(M-P)L}{N}
\end{align}
which can be written as,
\begin{align}
PL+\frac{(M-P)L}{N} &\leq \sum_{n=1}^N H(A_n^{[\cp]}|\cq)
\end{align}
which further can be written as,
\begin{align}
\left(1+\frac{M-P}{PN}\right)PL &\leq \sum_{n=1}^N H(A_n^{[\cp]}|\cq)
\end{align}
which leads to the desired converse result,
\begin{align}
\sum_{i=1}^PR_i=\frac{PL}{\sum_{n=1}^N H\left(A_n^{[\mathcal{P}]}\right)} \leq \frac{PL}{\sum_{n=1}^N H\left(A_n^{[\mathcal{P}]}|\cq\right)} \leq \frac{1}{1+\frac{M-P}{PN}}
\end{align}

\subsection{Converse Proof for the Case $\frac{M}{P} > 2$}\label{converse2}

In the sequel, we derive an inductive relation that can be used in addition to the base induction step of $1 \leq \frac{M}{P} \leq 2$ derived in the previous sub-section to obtain an upper bound for the MPIR problem. The idea we pursue here is similar in spirit to the one in \cite{JafarPIR}, where the authors developed a base converse step for $M=2$ messages, and developed an induction over the number of messages $M$ for the case $M>2$. Here, we have developed a base converse step for $1\leq \frac{M}{P}\leq 2$, and now develop an induction over $\left\lfloor\frac{M}{P}\right\rfloor$ for the case $\frac{M}{P} > 2$.

The following lemma upper bounds the remaining uncertainty of the answer strings after knowing a subset of size $P$ of the interference messages.

\begin{lemma}[Interference Conditioning Lemma]
The remaining uncertainty in the answer strings $A_{2:N}^{[\cp_2]}$ after conditioning on the messages indexed by $\cp_1$, such that $\cp_1 \cap \cp_2=\phi$, $|\cp_1|=|\cp_2|=P$ is upper bounded by,
\begin{align}
H(A_{2:N}^{[\cp_2]}|W_{\cp_1}, \cq) \leq (N-1)[NH(A_1|\cq)-PL]
\end{align}
\end{lemma}

\begin{Proof}
We begin with
\begin{align}
H(A_{2:N}^{[\cp_2]}&|W_{\cp_1}, \cq)  \notag \\
\label{cond1}&\leq \sum_{n=2}^N H(A_n^{[\cp_2]}|W_{\cp_1}, \cq) \\
\label{cond2}&\leq \sum_{n=2}^N H(A_{1:n-1}^{[\cp_1]},A_n^{[\cp_2]},A_{n+1:N}^{[\cp_1]}|W_{\cp_1},\cq) \\
&= \sum_{n=2}^N H(A_{1:n-1}^{[\cp_1]},A_n^{[\cp_2]},A_{n+1:N}^{[\cp_1]},W_{\cp_1}|\cq)-H(W_{\cp_1}|\cq) \\
\label{cond3}&= \sum_{n=2}^N H(A_{1:n-1}^{[\cp_1]},A_n^{[\cp_2]},A_{n+1:N}^{[\cp_1]}|\cq)+H(W_{\cp_1}|A_{1:n-1}^{[\cp_1]},A_n^{[\cp_2]},A_{n+1:N}^{[\cp_1]})-H(W_{\cp_1}) \\
\label{cond4}&\leq \sum_{n=2}^N NH(A_1|\cq)-H(W_{\cp_1}) \\
&=(N-1)[NH(A_1|\cq)-PL]
\end{align}
where (\ref{cond1}) follows from the independence bound; (\ref{cond2}) follows from the non-negativity of entropy; (\ref{cond3}) follows from the statistical independence between the messages and the queries; and (\ref{cond4}) follows from the decodability of $W_{\cp_1}$ given the answer strings $(A_{1:n-1}^{[\cp_1]},A_n^{[\cp_2]}, \break A_{n+1:N}^{[\cp_1]})$, which is tantamount to the privacy constraint as in the second simplification.
\end{Proof}

Now, we derive the inductive relation for $\frac{M}{P} > 2$. Without loss of generality, let $\cp_1=\{1, \cdots, P\}$ and $\cp_2=\{P+1, \cdots, 2P\}$. Then, starting from (\ref{c4}), we write
\begin{align}
ML=&H(A_1,A_{2:N}^{[\cp_1]},A_{2:N}^{[\cp_2]}, \cdots, A_{2:N}^{[\cp_\beta]}|\cq) \\
  =&H(A_1,A_{2:N}^{[\cp_1]}|\cq)+H(A_{2:N}^{[\cp_2]}|A_1,A_{2:N}^{[\cp_1]},\cq)+H(A_{2:N}^{[\cp_3]}, \cdots, A_{2:N}^{[\cp_\beta]}|A_1,A_{2:N}^{[\cp_1]},A_{2:N}^{[\cp_2]},\cq) \\
  	\label{ind1}  \leq& NH(A_1|\cq)+H(A_{2:N}^{[\cp_2]}|A_1,A_{2:N}^{[\cp_1]},W_{1:P},\cq)\notag\\
  &+H(A_{2:N}^{[\cp_3]}, \cdots, A_{2:N}^{[\cp_\beta]}|A_1,A_{2:N}^{[\cp_1]},A_{2:N}^{[\cp_2]},W_{1:2P},\cq) \\
  	\label{ind2} \leq & NH(A_1|\cq)+H(A_{2:N}^{[\cp_2]}|W_{1:P},\cq)+H(A_{2:N}^{[\cp_3]},\! \cdots\!, A_{2:N}^{[\cp_\beta]}|A_1,W_{1:2P},\cq) \\
  	=&NH(A_1|\cq)+H(A_{2:N}^{[\cp_2]}|W_{1:P},\cq)\!+\!H(A_{1:N}^{[\cp_3]},\! \cdots\!, A_{1:N}^{[\cp_\beta]}|W_{1:2P},\cq)\!-\!H(A_1|W_{1:2P},\cq) \\
\label{ind3} =& NH(A_1|\cq)+H(A_{2:N}^{[\cp_2]}|W_{1:P},\cq)+(M-2P)L-H(A_1|W_{1:2P},\cq) \\
\label{ind4} \leq& NH(A_1|\cq)+(N-1)[NH(A_1|\cq)-PL]+(M-2P)L-H(A_1|W_{1:2P},\cq)
\end{align}
where (\ref{ind1}) follows from the decodability of $W_{1:2P}$ given $(A_1,A_{2:N}^{[\cp_1]},A_{2:N}^{[\cp_2]})$, the symmetry lemma and the independence bound; (\ref{ind2}) follows from the fact that conditioning does not increase entropy. In (\ref{ind3}), we note that subsets $(\cp_3, \cdots, \cp_\beta)$ include all messages $(W_1, \cdots, W_M)$ because every message appears in $\binom{M-1}{P-1}$ subsets. Hence, $H(A_{1:N}^{[\cp_3]}, \cdots, A_{1:N}^{[\cp_\beta]}| \break W_{1:2P},\cq)=(M-2P)L$ since $W_{2P+1:M}$ is decodable from $(A_{1:N}^{[\cp_3]}, \cdots, A_{1:N}^{[\cp_\beta]})$ after knowing $W_{1:2P}$. Finally, (\ref{ind4}) follows from the interference conditioning lemma.

Consequently, (\ref{ind4}) can be written as
\begin{align}
N^2H(A_1|\cq) &\geq (N+1)PL+H(A_1|W_{1:2P},\cq)
\end{align}
which is equivalent to
\begin{align}
\label{inductive_relation} NH(A_1|\cq) &\geq \left(1+\frac{1}{N}\right)PL+\frac{1}{N}H(A_1|W_{1:2P},\cq)
\end{align}
Now, (\ref{inductive_relation}) constructs an inductive relation, since evaluating $NH(A_1|W_{1:2P},\cq)$ is the same as $NH(A_1|\cq)$ with $(M-2P)$ messages, i.e., the problem of MPIR with $M$ messages for fixed $P$ is reduced to an MPIR problem with $(M-2P)$ messages for the same fixed $P$. We note that (\ref{inductive_relation}) generalizes the inductive relation in \cite{JafarPIR} for $P=1$.

We can write the induction hypothesis for MPIR with $M$ messages as
\begin{align}
NH(A_1|\cq) \geq PL \left[\sum_{i=0}^{\lfloor \frac{M}{P}\rfloor-1} \frac{1}{N^i}+\left(\frac{M}{P}-\left\lfloor\frac{M}{P}\right\rfloor\right)\frac{1}{N^{\left\lfloor \frac{M}{P}\right\rfloor}}\right]
\end{align}

Next, we proceed with proving this relation for $M+1$ messages. From the induction hypothesis, we have
\begin{align}
NH(A_1|W_{1:2P},\cq) &\geq PL \left[\sum_{i=0}^{\lfloor \frac{M-2P+1}{P}\rfloor-1} \frac{1}{N^i}+\left(\frac{M-2P+1}{P}-\left\lfloor\frac{M-2P+1}{P}\right\rfloor\right)\frac{1}{N^{\left\lfloor \frac{M-2P+1}{P}\right\rfloor}}\right]\\
                     &=PL \left[\sum_{i=0}^{\lfloor \frac{M+1}{P}\rfloor-3} \frac{1}{N^i}+\left(\frac{M+1}{P}-\left\lfloor\frac{M+1}{P}\right\rfloor\right)\frac{1}{N^{\left\lfloor \frac{M+1}{P}\right\rfloor-2}}\right]
\end{align}
substituting this in (\ref{inductive_relation}),
\begin{align}
NH(A_1|\cq) &\geq \left(1+\frac{1}{N}\right)PL+\frac{PL}{N^2}\left[\sum_{i=0}^{\lfloor \frac{M+1}{P}\rfloor-3} \frac{1}{N^i}+\left(\frac{M+1}{P}-\left\lfloor\frac{M+1}{P}\right\rfloor\right)\frac{1}{N^{\left\lfloor \frac{M+1}{P}\right\rfloor-2}}\right]\\
           &=PL \left[\sum_{i=0}^{\lfloor \frac{M+1}{P}\rfloor-1} \frac{1}{N^i}+\left(\frac{M+1}{P}-\left\lfloor\frac{M+1}{P}\right\rfloor\right)\frac{1}{N^{\left\lfloor \frac{M+1}{P}\right\rfloor}}\right] \label{lastinduction}
\end{align}
which concludes the induction argument.

Consequently, the upper bound for the MPIR problem can be obtained as,
\begin{align}
\sum_{i=1}^P R_i &= \frac{PL}{\sum_{n=1}^N H\left(A_n^{[\mathcal{P}]}\right)}\\
                 &\leq \frac{PL}{NH(A_1|\cq)}\\
                 &=\frac{1}{\sum_{i=0}^{\lfloor \frac{M}{P}\rfloor-1} \frac{1}{N^i}+\left(\frac{M}{P}-\left\lfloor\frac{M}{P}\right\rfloor\right)\frac{1}{N^{\left\lfloor \frac{M}{P}\right\rfloor}}} \label{interstep}\\
                 &=\left(\frac{1-(\frac{1}{N})^{\lfloor \frac{M}{P}\rfloor}}{1-\frac{1}{N}}+\left(\frac{M}{P}-\left\lfloor\frac{M}{P}\right\rfloor\right)\frac{1}{N^{\left\lfloor \frac{M}{P}\right\rfloor}}\right)^{-1} \label{laststep}
\end{align}
where (\ref{interstep}) follows from (\ref{lastinduction}); and (\ref{laststep}) follows from evaluating the sum in (\ref{interstep}).

\section{Conclusions}

In this paper, we introduced the multi-message private information retrieval (MPIR) problem from an information-theoretic perspective. The problem generalizes the PIR problem in \cite{JafarPIR} which retrieves a single message privately. We determined the exact sum capacity for this problem when the number of desired messages is at least half of the number of total stored messages to be $C_s^P=\frac{1}{1+\frac{M-P}{PN}}$. We showed that joint retrieval of the desired messages strictly outperforms repeating the single-message capacity achieving scheme for each message. Furthermore, we showed that if the total number of messages is an integer multiple of the number of desired messages, then the sum capacity is $C_s^P=\frac{1-\frac{1}{N}}{1-(\frac{1}{N})^{M/P}}$, which resembles the single-message PIR capacity expression when the number of messages is $\frac{M}{P}$. For the remaining cases, we derived lower and upper bounds. We observed numerically that the gap between the lower and bounds decreases monotonically in $N$, and the worst case gap is $0.0082$ which occurs for the case $N=2$ when $M=5$, $P=2$.

\bibliographystyle{unsrt}
\bibliography{references}
\end{document}